%% file: apssamp.tex
\documentclass[
reprint,
amsmath,amssymb,
aps,
pra,
showkeys,
]{revtex4-2}

\usepackage{graphicx}
\usepackage{dcolumn}
\usepackage{bm}
\PassOptionsToPackage{hyphens}{url}

\usepackage{hyperref}
\hypersetup{colorlinks=true,linkcolor=black,anchorcolor=black,citecolor=black,filecolor=black,urlcolor=black}
\usepackage[utf8]{inputenc}
\usepackage{siunitx}
\usepackage[version=4]{mhchem}
\usepackage{makecell}
\usepackage[mathlines]{lineno}% Enable numbering of text and display math
\usepackage{diagbox}
\usepackage{orcidlink}
\usepackage[caption=false]{subfig}
\usepackage{longtable}
\usepackage{bm}
\usepackage[defaultcolor=red]{changes}
\begin{document}

\title{Review of MEMS {Transducers} for Audio Applications}

\author{Nils Wittek\orcidlink{0009-0003-9755-6338}}
	\email{nils.wittek@ifm.uni-stuttgart.de}
  	\affiliation{University of Stuttgart, Institute for Micro Integration (IFM), 70569 Stuttgart, Germany}
  	\affiliation{Bosch Sensortec GmbH, Robert-Bosch-Ring 1, 01109 Dresden, Germany}

\author{Anton Melnikov\orcidlink{0000-0001-9466-653X}}
 	\email{anton.melnikov@bosch-sensortec.com}
	\affiliation{Bosch Sensortec GmbH, Robert-Bosch-Ring 1, 01109 Dresden, Germany}

\author{Bert Kaiser\orcidlink{0000-0002-7676-1506}}
	\email{bert.kaiser@bosch-sensortec.com}
	\affiliation{Bosch Sensortec GmbH, Robert-Bosch-Ring 1, 01109 Dresden, Germany}

\author{André Zimmermann\orcidlink{0000-0003-1824-9376}}
	\email{andre.zimmermann@ifm.uni-stuttgart.de}
	\affiliation{University of Stuttgart, Institute for Micro Integration (IFM), 70569 Stuttgart, Germany}
	\affiliation{Hahn-Schickard, 70569 Stuttgart, Germany}

\date{\today}

\begin{abstract}
Microelectromechanical systems (MEMS) speakers are compact, scalable alternatives to traditional voice coil speakers, promising improved sound quality through precise semiconductor manufacturing. This review provides an overview of the research landscape, covering {baseband-displacement}, {ultrasound-based}, and {thermoacoustic} sound generation concepts, classifying MEMS speakers by their actuation principles as electrodynamic, piezoelectric, or electrostatic devices. A comparative analysis of performance indicators {from 1990 to 2026} highlights the dominance of piezoelectric MEMS with {baseband displacement}, focusing on miniaturization and efficiency. The review outlines upcoming research challenges and identifies potential candidates for achieving full-spectrum audio performance. A~focus on innovative approaches could lead to widespread adoption of MEMS-only speakers.

\end{abstract}

\maketitle

\section{Introduction}

Microelectromechanical systems (MEMS) are integral to modern technology. From inertial sensors in cars to accelerometers and microphones in smart consumer devices, they have widely replaced parts manufactured by precision engineering. Their miniaturization, scalability through cost-effective mass production, and energy efficiency have enabled the success of many products essential to today's life. However, MEMS speakers remain underrepresented, with smartphones and in-ear headphones still relying on century-old magnet-based technology~\cite{3_zentner_mems_2021}.

\begin{figure*}[hbt!]
	\centering
	\subfloat[]{
		\begin{minipage}[b]{0.45\linewidth}
			\includegraphics[width=\columnwidth]{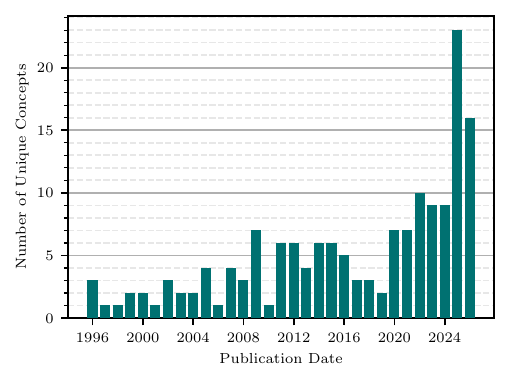}
			\label{fig:number_of_concepts_over_years}
			\vspace*{-1.5em}
		\end{minipage}
	}
	\hfill
	\subfloat[]{
		\begin{minipage}[b]{0.45\linewidth}
			\includegraphics[width=\columnwidth]{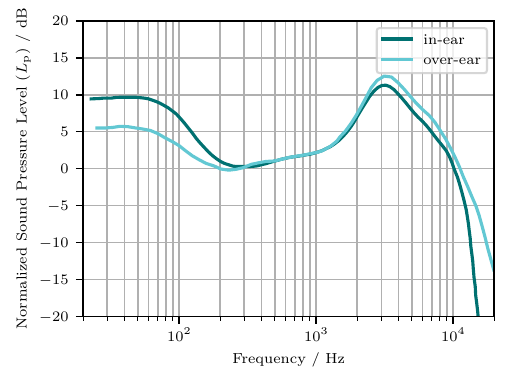}
			\label{fig:target_curve}
			\vspace*{-1.5em}
		\end{minipage}
	}
	\vspace{0.5cm}
	\subfloat[]{
		\begin{minipage}[b]{0.3\linewidth}
			\def\svgwidth{\columnwidth}
			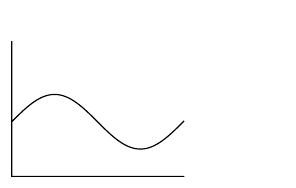
			\label{fig:DSR_principle}
		\end{minipage}
	}
	\hfill
	\subfloat[]{
		\begin{minipage}[b]{0.3\linewidth}
			\def\svgwidth{\columnwidth}
			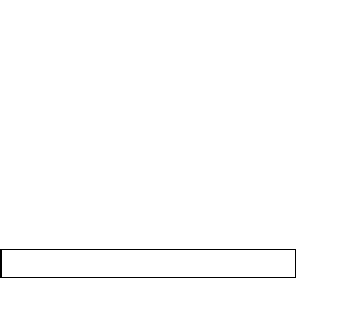
			\label{fig:mass-spring-oscillator}
		\end{minipage}
	}
	\hfill
		\subfloat[]{
		\begin{minipage}[b]{0.3\linewidth}
			\def\svgwidth{\columnwidth}
			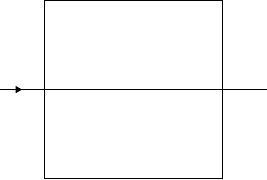
			\label{fig:lorentz_force}
		\end{minipage}
	}
	\vspace{0.5cm}
	\subfloat[]{
		\begin{minipage}[b]{0.3\linewidth}
			\def\svgwidth{\columnwidth}
			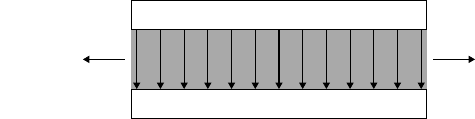
			\label{fig:piezoelectric_drive}
		\end{minipage}
	}
	\hfill
	\subfloat[]{
		\begin{minipage}[b]{0.3\linewidth}
			\def\svgwidth{\columnwidth}
			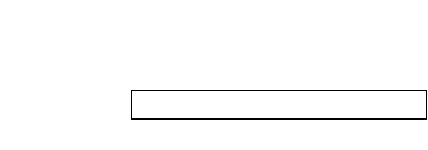
			\label{fig:pull-configuration}
		\end{minipage}
	}
	\hfill
	\subfloat[]{
		\begin{minipage}[b]{0.3\linewidth}
			\def\svgwidth{\columnwidth}
			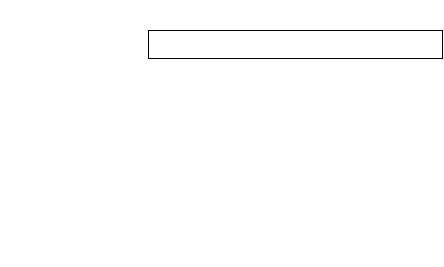
			\label{fig:push-pull-configuration}
		\end{minipage}
	}
	\caption[]{\textbf{Overview of MEMS speaker publication trends and schematic illustrations of different concepts.}
		\subref{fig:number_of_concepts_over_years}~{Combined number of journal articles and conference papers on unique MEMS audio speaker concepts published per year from 1990 to May 2026.}
		\subref{fig:target_curve}~Harman target curve for in-ear and over-ear headphones~\cite{olive_preferred_2016}.
		\subref{fig:DSR_principle}~Concept of digital sound reconstruction (DSR) and depiction of an ideal sound pulse on the right, adapted from~\cite{17_diamond_digital_2003,4_mayrhofer_new_2021}.
		\subref{fig:mass-spring-oscillator}~Schematic of the base components of a speaker, comprising mass~$m$, damper~{$b$}, and spring~{$k$}. Drive force~$F$ acts in the $x$ direction.		
		\subref{fig:lorentz_force}~Lorentz force~$\vec{F}_{\mathrm{mag}}$ acting on a conductor with length~$\vec{l}$ carrying a current~$i$ perpendicular to an external magnetic field~$\vec{B}$.
		\subref{fig:piezoelectric_drive}~Signal voltage~{$U_{\mathrm{AC}}$} applied across two electrodes with a piezoelectric material in between, resulting in strain~{$e$} due to the inverse piezoelectric effect in the~$d_{31}$ direction.
		\subref{fig:pull-configuration}~Schematic of an electrostatic actuator in pull configuration, with the movable top electrode displaced towards the bottom electrode because of the electrostatic force~$F_{\mathrm{el}}$ upon application of the signal voltage~{$U_{\mathrm{AC}}$}.
		\subref{fig:push-pull-configuration}~Schematic of an electrostatic actuator in push-pull configuration with signal voltage~{$U_{\mathrm{AC}}$} applied on the movable center electrode and bias voltages of~{$\pm U_{\mathrm{DC}}$} on the fixed electrodes.}
\end{figure*}

Since the 1960s, loudspeaker technology has seen limited innovation, continuing to use principles from the late 19th century~\cite{visse_microphones_2025}. Advances in semiconductor manufacturing since the early 2000s have sparked research on MEMS-based speaker designs~\cite{visse_microphones_2025}. This growing research activity is reflected in the increasing number of {published MEMS audio speaker concepts per year}, as shown in Fig.~\ref{fig:number_of_concepts_over_years}. From a market perspective, MEMS speakers are entering a phase of rapid innovation, while electrodynamic speakers have reached saturation~\cite{visse_microphones_2025}, indicating a potential disruptive shift in line with Christensen's theory of disruptive innovation~\cite{christensen_innovators_1997}.

According to Yole market research~\cite{visse_microphones_2025}, MEMS speakers are expected to enter the market through true wireless stereo (TWS) devices, where their compact size and energy efficiency are critical for battery life and form factor. However, TWS systems often include active noise control, which imposes additional performance requirements~\cite{kuo_active_1999}.

Listeners typically prefer specific frequency responses when listening to music, leading to standardized target curves such as the Harman curve shown in Fig.~\ref{fig:target_curve} for in-ear and over-ear headphones~\cite{olive_preferred_2016}. Each speaker has a unique frequency response, requiring sufficient sound pressure levels (SPL) from the loudspeaker in amplified regions and attenuation from the entire system in others. Therefore, the speaker has to produce sounds with at least the same SPL as ambient noise, with the ability to reproduce audio playback on top of it. Especially at low frequencies, speakers must reach SPLs of up to $\SI{140}{\dB}$, also accounting for leakage losses~\cite{14_moore_speaker_2023,42_noauthor_sound_2024}.

Building on recent research, three review articles have previously addressed MEMS speakers for audio applications. Garud et al.~\cite{garud_mems_2024} discuss the fundamental drive mechanisms, research progress in India, and commercial developments. Gemelli et al.~\cite{gemelli_recent_2023} review MEMS speakers alongside microphones, covering drive and sensing principles as well as interface circuits. Wang et al.~\cite{52_wang_review_2021} focus on drive concepts and fabrication methods, highlighting potential performance improvements but excluding {ultrasound-based} designs (as defined in Sec.~\ref{sec:ultrasound}). Although not providing a review, Sun et al.~\cite{176_sun_figure_2025} introduce two performance indicators that simplify the design and optimization of piezoelectric MEMS cantilever speakers across the full frequency range.

This paper provides an international overview of MEMS speaker development for audio applications, with emphasis on recent progress. It presents a detailed analysis of scientific research, including aggregated performance data for qualitative and quantitative comparison. By analytically converting results from different measurement setups, this study enables cross-comparison of speakers with fundamentally different working principles. This supports a deeper understanding of sound generation concepts, including {ultrasound-based} approaches, while highlighting key design drivers and helping identify promising future concepts. {Detailed considerations of driving circuits, system-level design, power consumption, and distortion are beyond the scope of this article and discussed in~\cite{gemelli_recent_2023}.}

{The article first establishes the foundational acoustic parameters, sound generation concepts, and drive principles of MEMS speakers. Building on these fundamentals, we systematically review and discuss the diverse research approaches in the field. Finally, an analysis of literature data is presented, including performance trade-offs and technological trends.}

\section{Acoustic Parameters and Measurement Methods}\label{sec:acoustic_parameters}
To enable comparison across diverse MEMS speaker designs, this section introduces key acoustic parameters and outlines standardized measurement methodologies.

\subsection{Acoustic Parameters}

A loudspeaker produces sound by displacing a volume of air~$V$. Under free-field conditions, {the generation of propagating pressure waves~$p$ requires volume acceleration}. In a small, sealed volume like an ear canal, the displacement directly produces pressure changes~$p$ of the enclosed air. The acoustic performance is quantified by the sound pressure level (SPL), defined in Eq.~\eqref{eq:SPL}, which relates the root-mean-square pressure~$p_{\mathrm{rms}}$ to the reference pressure~$p_{\mathrm{ref}} = \SI{20}{\micro\Pa}$, the approximate threshold of human hearing, expressed in decibels~($\si{\dB}$)~\cite{IEC60268-21}. In the context of this article, all SPLs are reported in~\si{\decibel} with a reference pressure of~\SI{20}{\micro\pascal}.

\begin{equation}\label{eq:SPL}
	L_\mathrm{p} = 20 \cdot \log_{10} \left( \frac{p_{\mathrm{rms}}}{p_{\mathrm{ref}}} \right) \si{\decibel}
\end{equation}

Total harmonic distortion (THD) is a key metric for assessing loudspeaker signal quality. It quantifies deviations from a pure sine wave when the speaker is driven by an undistorted input. The most common definition in audio, given in Eq.~\eqref{eq:THD}, expresses the ratio of higher-order harmonic pressure amplitudes~$\sum \hat{p}_n^2 - \hat{p}_1^2$ to the fundamental~$\hat{p}_1^2$. Since this ratio can exceed unity, an alternative formulation normalizes by the total harmonic content~\cite{IEC60268-21}.

\begin{equation}\label{eq:THD}
	THD = 100 \si{\percent} \cdot \sqrt{\frac{\sum \hat{p}_n^2 - \hat{p}_1^2}{\hat{p}_1^2}}
\end{equation}

Both SPL and THD are usually collected over a frequency range, as both parameters are frequency dependent. {For comprehensive device characterization, two primary measurement approaches are employed.} The first approach is to measure the sound generation of the speaker into a closed chamber, which is most often represented by a standardized ear coupler~\cite{3_zentner_mems_2021,IEC60318-4}. The second method is to measure under free-field conditions with a microphone placed at a defined distance from the loudspeaker~\cite{IEC60268-21,IEC60268-22}.

These parameters also influence perceived audio quality. Although dedicated listener studies on MEMS speakers are lacking, tight fabrication tolerances improve device uniformity and multi‑speaker reproduction. {A balanced ear‑coupler response} simplifies target tuning, and low moving mass enables fast transient response, all of which are expected to enhance perceived sound quality.~\cite{borwick_loud_2001}

\subsection{Ear Coupler Measurements}
{When an earbud is inserted into the ear canal, it forms a sealed cavity.} The motion of the speaker diaphragm induces pressure variations within this chamber. Assuming adiabatic conditions, linearization of the adiabatic equation from Ref.~\cite{karaoglu_classical_2020} through a Taylor expansion yields Eq.~\eqref{eq:linearized_adiabatic_equation}, which relates {the displaced volume amplitude~$\hat{V}$ to the pressure amplitude~$\hat{p}$}. Here, $\kappa$ is the adiabatic constant of air, $p_0$ the atmospheric pressure, and $V_0$ the initial volume of the chamber. {According to IEC 60318-4~\cite{IEC60318-4}, the initial volume of a standardized ear coupler is $V_0 = \SI{1.26}{\cubic\cm}$.}

\begin{align}
	\hat{p} &= - \frac{\kappa \, p_0}{V_0} \hat{V} \label{eq:linearized_adiabatic_equation}
\end{align}

Inserting Eq.~\eqref{eq:linearized_adiabatic_equation} into Eq.~\eqref{eq:SPL} yields an approximation for the SPL resulting from a given volume displacement, as shown in Eq.~\eqref{eq:SPL_adiabatic}. The factor~$1/\sqrt{2}$ accounts for the conversion from pressure amplitude to root-mean-square pressure for sinusoidal signals. {This quasi-static approach is valid for device diameters below the acoustic wavelength~\cite{borwick_loud_2001}, corresponding to $d < \SI{110}{\milli\meter}$ for an acoustic frequency of~\SI{1}{\kilo\hertz}.} {All speakers reviewed in this article satisfy this size criterion, with most being significantly smaller.} The approach does not capture resonance effects introduced by the ear coupler geometry.

\begin{equation}\label{eq:SPL_adiabatic}
	L_\mathrm{p} = 20 \, \log_{10} \left( \frac{\kappa \, p_0}{\sqrt{2} \, V_0 \, p_{\mathrm{ref}}} \hat{V} \right) \si{\decibel}
\end{equation}

{Ear couplers reported in the literature vary in volume, requiring conversion for consistent comparison.} SPL values can be converted to the equivalent SPL in a standard ear coupler according to IEC~60318-4~\cite{IEC60318-4} using Eq.~\eqref{eq:conversion_coupler_volumes}. In this context, $V_1$ is the volume of the measurement coupler, and~$L_{\mathrm{p}_1}$ the corresponding SPL. The conversion equation is derived by solving Eq.~\eqref{eq:SPL_adiabatic} for~$\hat{V}$ using~$L_\mathrm{p} = L_{\mathrm{p}_1}$, and substituting the result back into Eq.~\eqref{eq:SPL_adiabatic} with~$L_\mathrm{p} = L_{\mathrm{p}_0}$.

\begin{equation}\label{eq:conversion_coupler_volumes}
	L_{\mathrm{p}_0} = 20 \cdot \log_{10} \left( \frac{10^{\frac{L_{\mathrm{p}_1}}{20}} \, V_1}{V_0} \right) \si{\decibel}
\end{equation}

\subsection{Free-Field Measurements}
In free-field measurements, a microphone is positioned at a known distance from the speaker, which is typically mounted on a standardized quasi-infinite baffle to prevent acoustic shorts. The speaker radiates sound omnidirectional into half-space, {as long as it can be considered as a monopole}~\cite{IEC60268-21,IEC60268-22}. In this case, the radiated pressure $p$ is given by Eq.~\eqref{eq:pressure_point_source}, where $\rho$ is the air density and {$K$} the wave number. The volume flow is denoted by {$q = \dot{V}$}, the distance from the source by~$r$, and~i is the imaginary unit~\cite{williams_fourier_1999}. This model neglects directivity, wavelengths comparable to the device dimensions, and high-frequency behavior.

\begin{equation}\label{eq:pressure_point_source}
	p = \frac{- 2 \, \mathrm{i} \, \rho \, {v} \, {K}}{4 \pi} \, {q} \, \frac{e^{\mathrm{i} \, {K} \, r}}{r}
\end{equation}

By substituting {$K = \frac{2 \pi f}{v}$} in Eq.~\eqref{eq:pressure_point_source}, where $f$~is the oscillation frequency, solving for {$q$}, and integrating over time, the displaced volume amplitude~$\hat{V}$ is obtained as shown in Eq.~\eqref{eq:volume_amplitude_free-field}. The pressure amplitude~{$\hat{p}$} of the radiated wave is given by solving Eq.~\eqref{eq:volume_amplitude_free-field} for~$\hat{p}$. {It is important to note that this approach is only valid under quasi-static conditions, for device diameters below the acoustic wavelength~\cite{borwick_loud_2001}.}

\begin{equation}\label{eq:volume_amplitude_free-field}
	\hat{V} = \frac{r \, \hat{p}}{\rho \, 2 \pi f^2}
\end{equation}

{The SPL decreases by approximately~\SI{6}{\decibel} when the distance between source and receiver is doubled~\cite{borwick_loud_2001}. To compare the SPL measured at different distances under free-field conditions, Eq.~\eqref{eq:distance_free-field} can be used to calculate the SPL at distance~{$r_2$} from a known value at~{$r_1$}. Here,~{$L_{\mathrm{p}_{\mathrm{r}_1}}$} and~{$L_{\mathrm{p}_{\mathrm{r}_2}}$} denote the SPLs at~{$r_1$} and~{$r_2$}, respectively.}

\begin{equation}\label{eq:distance_free-field}
	{L_{\mathrm{p}_{\mathrm{r}_2}} = L_{\mathrm{p}_{\mathrm{r}_1}} - 20 \cdot \log_{10} \left( \frac{r_2}{r_1} \right) \si{\decibel}}
\end{equation}

\subsection{Comparability between Ear Coupler and Free-Field Measurements}
{A direct conversion between in-ear and free-field measurements is not trivial as the underlying acoustic principles are distinct. A comparison is only justifiable under quasi-static conditions which account for the different relationship between acoustic pressure~$\hat{p}$ and volume displacement~$\hat{V}$ in each regime. Ideally, speakers should be characterized using both standardized measurement methods to enable comprehensive comparison. However, {testing is often limited} to a single method, typically selected to reflect the intended application. The primary design trade-offs are governed by the acoustic boundary conditions, particularly whether the application operates under free-field conditions or within a closed cavity. For in-ear headphones, small size and low power consumption are particularly important. Free-field applications can accommodate larger devices and higher power consumption but impose more demanding acoustic performance requirements.}

{Assuming a constant volume displacement source,} the resulting pressure~$\hat{p}$ in an ear coupler pressure chamber is proportional to~$\hat{V}$ at low frequencies, as shown in Eq.~\eqref{eq:linearized_adiabatic_equation}. In contrast, when the same source 
radiates under free-field conditions, $\hat{p}$~is proportional to volume acceleration. This is represented by the $f^2$~dependence in Eq.~\eqref{eq:volume_amplitude_free-field}, which relates free-field pressure to displacement ($\hat{p} \propto f^2 \, \hat{V}$). Therefore, the conversion from a free-field SPL~$L_{\mathrm{p}_{\mathrm{ff}}}$ to an equivalent ear coupler SPL~$L_{\mathrm{p}_{\mathrm{ec}}}$ as given by Eq.~\eqref{eq:equivalent_SPL} provides a frequency-dependent approximation rather than a general equivalence, enabling comparison when only one measurement type is available. The equation is derived by solving Eq.~\eqref{eq:SPL} for~$p_{\mathrm{rms}}$, which is converted to a pressure amplitude. This value is then inserted into Eq.~\eqref{eq:volume_amplitude_free-field}, and the result used in Eq.~\eqref{eq:SPL_adiabatic} for the final derivation. This conversion remains {an approximation, valid} in the quasi-static {range below the electro-mechanical resonance of the loudspeaker and neglects high-frequency directivity effects} where the monopole assumption is invalid.

\begin{align}
	L_{\mathrm{p}_{\mathrm{ec}}} &= 20 \cdot \log_{10} \left( \frac{\kappa \, p_0 \, r \, 10^{\frac{L_{\mathrm{p}_{\mathrm{ff}}}}{20}}}{V_0 \, \rho \, 2 \pi f^2} \right) \si{\decibel} \label{eq:equivalent_SPL}
\end{align}

\section{Sound Generation Concepts}
{To systematically categorize the principal sound generation concepts in MEMS speaker research, transducers are primarily classified by whether they employ analog sound reproduction or digital sound reconstruction (DSR). Within the analog domain, sound generation mechanisms are distinguished by whether pressure waves are generated through the physical displacement of a solid boundary or via periodic temperature fluctuations in the surrounding medium using the thermoacoustic effect. Displacement-based transducers are further categorized by their primary operating frequency regime: baseband-displacement speakers, which actuate directly within the audible spectrum, and carrier-based speakers, which modulate an ultrasonic carrier wave and rely on subsequent demodulation. Alongside these carrier-based systems, DSR constitutes the other major ultrasound-based approach. This classification establishes three primary categories in modern MEMS design, introduced in the following subsections.}

\subsection{{Baseband Displacement}}
{Baseband displacement} is the most established method of sound generation in loudspeakers, originating with Alexander Graham Bell’s 1876 telephone patent. It describes not only drivers radiating sound into an open environment but also those operating within a closed chamber, where the primary function is to generate acoustic pressure changes. {Baseband-displacement} speakers generate sound through the motion of a body, usually a diaphragm {oscillating at the audio frequency}. In audio terminology, this movement is referred to as the excursion of the membrane.
As the diaphragm moves outward, it compresses air, increasing local density. When it {moves backward}, regions of lower air density are created. These pressure variations propagate as longitudinal acoustic waves {in a free-field environment}. Upon reaching the ear, they cause the eardrum to oscillate, enabling sound perception. From Eq.~\eqref{eq:volume_amplitude_free-field}, the pressure amplitude~$\hat{p}$ radiated by a {baseband-displacement} speaker, proportional to the displaced volume amplitude~$\hat{V}$, can be expressed by Eq.~\eqref{eq:direct_displacement}, where the effective diaphragm area~$A_{\mathrm{eff}}$ and displacement amplitude~$\hat{x}$ determine the radiated pressure. {This relationship is valid for a single frequency, while the frequency-dependent behavior can be found in the previous section}. The frequency of diaphragm motion sets the frequency of the emitted sound.~\cite{borwick_loud_2001}

\begin{equation}\label{eq:direct_displacement}
	\hat{p} \propto \hat{V} \propto \hat{x} \, A_{\mathrm{eff}}
\end{equation}

\subsection{{Ultrasound-Based}}\label{sec:ultrasound}
In contrast to conventional {baseband displacement}, {ultrasound-based} speakers operate with mechanically displaceable elements driven at ultrasonic frequencies ($>\SI{20}{\kilo\hertz}$), typically excited near their mechanical resonance~\cite{mayrhofer_mems_2024}. {They either digitally reconstruct the audible sound signal or modulate an ultrasonic carrier signal to generate the audible baseband signal through demodulation.} {While carrier-based speakers theoretically include parametric speakers, they remain outside the scope of this review.} Parametric speakers employ arrays of ultrasonic transducers to emit a modulated carrier signal that self-demodulates in air, thereby producing highly directional sound beams~\cite{shi_development_2010}.

Digital sound reconstruction (DSR) utilizes an array of micro-speakers, called speaklets. These devices are assigned to bit groups and represent digital values by switching between displaced (high) and non-displaced (low) states. Superposition of individual pulses creates an analog audio signal, also schematically shown in Fig.~\ref{fig:DSR_principle}. Unlike conventional speakers, sound amplitude is controlled by the number of actuated speaklets rather than diaphragm displacement~\cite{17_diamond_digital_2003}. Advanced digital sound reconstruction (ADSR) extends DSR by integrating shutters that allow speaklets to move without contributing to the output signal, effectively using phase-shifted pumps instead of many diaphragms~\cite{4_mayrhofer_new_2021}.

Carrier-based pump speakers generate sound by {displacing} air with a diaphragm oscillating at an ultrasonic carrier frequency~$f_{\mathrm{c}}$, where the oscillation is modulated by an audio signal to produce a time-varying volume flow into a cavity. The air is released on the opposite side through a second diaphragm, also driven at~$f_{\mathrm{c}}$, which acts as a shutter and demodulates the signal via variation of the {acoustic} resistance. The modulated diaphragm is typically driven using double-sideband suppressed-carrier (DSB-SC) or single-sideband suppressed-carrier (SSB-SC) modulation. {At any given instant}~$t$, modulation is achieved by multiplying the audio signal~$a(t)$ with the carrier~$c(t)$, typically a cosine function, as shown in Eq.~\eqref{eq:carrier}. For DSB-SC, this yields the modulated signal in Eq.~\eqref{eq:modulation}. In SSB-SC, one of the sidebands is removed, leading to a modulation of the frequency. In both cases, demodulation is performed by the second diaphragm, which is driven by~$c(t)$ only and effectively applies a second multiplication with the carrier, resulting in Eq.~\eqref{eq:demodulation} for DSB-SC. {The audio component~$a(t)$ is audible, while high-frequency components remain inaudible.}~\cite{mayrhofer_mems_2024,10_chen_modulated_2025}

\begin{align}
	c(t) &= \cos (2 \pi f_{\mathrm{c}} t) \label{eq:carrier} \\[6pt]
	s(t) &= \cos (2 \pi f_{\mathrm{a}} t) \, \cos (2 \pi f_{\mathrm{c}} t) \nonumber \\[6pt]
	&= \frac{1}{2} \left( \cos (2 \pi t (f_{\mathrm{a}} - f_{\mathrm{c}})) + \cos (2 \pi t (f_{\mathrm{a}} + f_{\mathrm{c}})) \right) \label{eq:modulation} \\[6pt]
	s(t) \, c(t) &= a(t) \, c(t)^2 = \frac{1}{2} a(t) \left( 1 + \cos (4 \pi f_{\mathrm{c}} t) \right) \label{eq:demodulation}
\end{align}

A variation of the pump-based concept uses a cavity with dimensions matched to a fraction of the carrier wavelength to form a resonance chamber. Amplitude-modulated diaphragms generate standing waves within the cavity. At points of maximum amplitude, additional diaphragms open the chamber to the opposite side, effectively demodulating the signal by releasing air at peak pressure locations.~\cite{liang_air-pulse_2024}

All mentioned pump-based approaches increase the SPL at low frequencies, as multiple pump cycles can occur within one period of the audio signal. However, this also leads to decreasing SPL with increasing frequency, following a~$\frac{1}{f}$ behavior. {Therefore, higher carrier frequencies are beneficial as they enable significant oversampling without limiting the achievable audio bandwidth.}~\cite{mayrhofer_mems_2024,10_chen_modulated_2025,146_tenorio_physics-based_2025}

Different modulation schemes yield varying ultrasonic energy content, which is highest for SSB-SC and lower for DSB-SC, while ADSR does not introduce ultrasonic harmonics. These harmonics can cause audible distortion when demodulated into the audio spectrum, and are primarily generated by nonlinear slit impedance and the coupling of diaphragms via pressure build-up in the shared cavity. The amplitude of these harmonic components is dependent on the diaphragm displacement, and the resulting distortion can be managed by the choice of modulation scheme.~\cite{mayrhofer_mems_2024} It is noted that other systemic nonlinearities, such as those from the mechanical, fluidic, and electric domains, may not degrade the audible signal provided that they are {time-invariant} and uniform across all elements in an array.

\subsection{{Thermoacoustic}}
As an alternative to mechanical displacement, {thermoacoustic} speakers generate sound via the thermoacoustic effect. The sound signal passes as an alternating current through a conductive element, inducing temperature fluctuations in the surrounding {medium}. These thermal changes cause expansion and contraction of the {medium}, producing pressure waves~\cite{mayo_advancements_2018}.

Several analytical models describe this phenomenon~\cite{arnold_thermophone_1917,wente_thermophone_1922,137_xiao_flexible_2008,hu_model_2010,vesterinen_fundamental_2010,23_tian_graphene--paper_2011}, focusing on frequency-dependent behavior and the influence of material properties. An important performance metric is the heat capacity per unit area, as lower values and larger surface areas enhance heat transfer to the surrounding medium~\cite{mayo_advancements_2018}.

Assuming negligible heat capacity and modeling the source as a {monopole}, the radiated pressure~$p$ is given by Eq.~\eqref{eq:thermoacoustic}. This expression depends on the sound frequency~$f$, electrical input power~$P_{\mathrm{el}}$, radial distance from the sound source~$r$, specific heat capacity of the surrounding medium~$c_{\mathrm{p}}$, and ambient temperature~$T_{\mathrm{amb}}$. Under these ideal conditions, speaker performance is governed solely by the properties of the surrounding medium~\cite{mayo_advancements_2018}.

\begin{equation}
	p = \frac{f \, P_{\mathrm{el}}}{\sqrt{2} \, r \, c_{\mathrm{p}} \, T_{\mathrm{amb}}} \label{eq:thermoacoustic}
\end{equation}

{As sound generation via the thermoacoustic effect depends on electrical input power~$P_{\mathrm{el}}$, temperature fluctuations and thus acoustic output occur at twice the frequency of the electrical AC input signal, because electrical power is proportional to the square of the voltage.} This effect can be mitigated by applying a DC bias~\cite{137_xiao_flexible_2008,141_niskanen_suspended_2009}.

\section{Drive Principles}
The mechanical behavior of a {displacement} speaker can, in its simplest form, be modeled by the differential equation of a damped harmonic oscillator, as shown in Eq.~\eqref{eq:ODE_mechanical}. This system comprises the components shown in Fig.~\ref{fig:mass-spring-oscillator}, with the moving mass~$m$, {mechanical} spring stiffness~{$k_\mathrm{m}$}, and damping coefficient~{$b$, which represents energy loss}. {The displacement of the mass, a function of time~$t$, is described by~$x$. Its first and second time derivatives are denoted by~$\dot{x} = dx/dt$ and~$\ddot{x} = d^2x/dt^2$, respectively.} The external force~$F$, which varies with the actuation method, distinguishes different drive principles.

\begin{equation}\label{eq:ODE_mechanical}
	m \ddot{x} + {b} \dot{x} + {k_\mathrm{m}} x = F
\end{equation}

\subsection{Electrodynamic}
When a conductor of length~$\vec{l}$ carrying a current~$i$ is placed in an external magnetic field~$\vec{B}$, the Lorentz force~$\vec{F}_{\mathrm{mag}}$ acts on the conductor, as defined in Eq.~\eqref{eq:lorentz_force}~\cite{di_barba_mems_2020}. {Following the standard convention for the Lorentz force in a wire~\cite{karaoglu_classical_2020}, the vector~$d\vec{l}$ represents a differential segment of the wire pointing in the direction of the scalar current~$i$.} If the current flow is oriented perpendicular to the magnetic field, the expression simplifies to a scalar form~\cite{lenk_electromechanical_2011} as illustrated in Fig.~\ref{fig:lorentz_force}.

\begin{align}
	\vec{F}_{\mathrm{mag}} &= i \left( \vec{l} \times \vec{B} \right) \label{eq:lorentz_force}
\end{align}

Regarding magnet-based actuation, the electrodynamic principle dominates MEMS speaker applications. Another principle used is magnetostriction, where a magnetostrictive layer expands or contracts under an external magnetic field~\cite{138_s_albach_sound_2011}. Two alternative principles are piezomagnetic and electromagnetic actuation, both currently not applied in MEMS speakers, and therefore excluded from this review. Further details on these principles are available in Ref.~\cite{lenk_electromechanical_2011}.

{Nonlinearities in electrodynamic speakers arise from magnetic field inhomogeneities~\cite{77_cheng_silicon_2004,127_lee_analysis_2010} and iron-induced effects, including eddy currents and magnetic reluctance, which can be mitigated by ironless magnetic circuits~\cite{90_lemarquand_mems_2012}.}

\subsection{Piezoelectric}
Piezoelectric materials generate electric charges under mechanical stress (direct piezoelectric effect) and deform under an applied electric field (inverse piezoelectric effect). The latter is relevant for MEMS speaker actuation. Both effects originate from the non-centrosymmetric charge distribution in the crystal lattice of natural piezoelectric materials like quartz crystals. An applied force leads to charge redistribution in the crystal unit cell, resulting in electrical dipole formation, referred to as polarization.~\cite{kempe_inertial_2011}

{Single-crystal piezoelectric materials are rarely used in MEMS due to integration challenges. Instead, polycrystalline thin-films including ferroelectric materials like lead zirconate titanate (PZT) are employed.} As polar materials, they exhibit spontaneous polarization below their Curie temperature in the form of Weiss domains with aligned dipoles. However, the random orientation of these domains results in zero net polarization. Therefore, an external electric field is applied during a poling process below the Curie temperature of the material to align the domains and induce macroscopic piezoelectric properties. {Due to the hysteretic nature of the poling process, the level of remanent polarization determines the effective, macroscopic piezoelectric performance of the fabricated material.}~\cite{kempe_inertial_2011}

After {poling}, piezoelectric materials can be described using a linear approximation by coupled Maxwell–Hooke relations, given in Eqs.~\eqref{eq:direct_piezoeffect} and \eqref{eq:inverse_piezoeffect} for the direct piezoelectric effect and inverse piezoelectric effect, respectively. Here, $\bar{D}$~denotes the electric displacement vector, $\bar{E}$~the electric field vector, and $\bar{e}$~and $\bar{\sigma}$~the strain and stress vectors in Voigt notation. The electrical permittivity matrix is~$\epsilon$, while superscripts indicate evaluation at zero or constant~$\bar{\sigma}$ or~$\bar{E}$. The matrix~$\mathbf{d}$ contains the direct piezoelectric coefficients or piezoelectric charge constants, relating stress to polarization and electric field to strain, and $\mathbf{s}$~denotes the elastic compliance matrix.~\cite{kempe_inertial_2011}

\begin{align}
	{\bar{D}} &{= \epsilon^{\bar{\sigma}} \bar{E} + \mathbf{d} \bar{\sigma}} \label{eq:direct_piezoeffect} \\[6pt]
	{\bar{e}} &{= \mathbf{s}^{\bar{E}} \bar{\sigma} + \mathbf{d}^{\mathrm{T}} \bar{E}} \label{eq:inverse_piezoeffect}
\end{align}

{For ferroelectric materials of the ditetragonal- or dihexagonal-pyramidal class (e.g., PZT), the relations simplify to the form given in Eq.~\eqref{eq:inverse_piezoeffect_matrix} for the inverse piezoelectric effect, assuming no external mechanical stress ($\bar{\sigma} = \mathbf{0}$).}~\cite{kempe_inertial_2011}

\begin{align}
	{\begin{bmatrix} e_1 \\ e_2 \\ e_3 \\ e_4 \\ e_5 \\ e_6 \end{bmatrix} }
	& {= 
	\begin{bmatrix} 0 & 0 & d_{31} \\ 0 & 0 & d_{32} \\ 0 & 0 & d_{33} \\ 0 & d_{24} & 0 \\ d_{15} & 0 & 0 \\ 0 & 0 & 0 \end{bmatrix} 
	\begin{bmatrix} E_1 \\ E_2 \\ E_3 \end{bmatrix}} \label{eq:inverse_piezoeffect_matrix}
\end{align}

{Directions are labeled 1, 2, and~3, corresponding to the $x$, $y$, and $z$~axes, with direction~3 typically aligned with polarization. For MEMS devices, the $d_{31}$~mode is most relevant, as thin-film piezoelectric layers provide good manufacturability and are typically poled in the $z$-direction~\cite{kempe_inertial_2011}. In this mode, the material expands perpendicular to the electrodes, as shown in Fig.~\ref{fig:piezoelectric_drive}.}

{After the prior simplifications, the inverse piezoelectric effect in this direction can be described by Eq.~\eqref{eq:strain_reverse_piezo_effect}, where the electric field~$E_3$ leads to a strain~$e_1$, with~$d_{31}$ being the piezoelectric charge constant. Using Eq.~\eqref{eq:stress_strain}, which relates mechanical stress~$\sigma$ and strain~$e$, and $\sigma = \frac{F}{A}$, the resulting force can be expressed as in Eq.~\eqref{eq:piezo_force}, with~$E = \frac{U}{g}$. Here, $s$~is the compliance, $A$~the electrode-facing surface area of the piezoelectric material, $U_{\mathrm{AC}}$~the signal voltage amplitude of the audio signal, and $g$~the material thickness between the electrodes.}~\cite{lenk_electromechanical_2011}

\begin{align}
	{e_1} &{= d_{31} \, E_3} \label{eq:strain_reverse_piezo_effect} \\[6pt]
	{\sigma} &{= \frac{e_1}{s} = \frac{d_{31} \, E_3}{s} = \frac{F}{A}} \label{eq:stress_strain} \\[6pt]
	{F} &{= \frac{d_{31} \, U_{\mathrm{AC}} \, A}{s \, g}} \label{eq:piezo_force}
\end{align}

The piezoelectric coefficients~$d_{ij}$ vary significantly depending on doping materials and fabrication processes~\cite{kempe_inertial_2011}. {When a piezoelectric thin film is combined with a cantilever, it creates a layered structure known as a unimorph or, when two piezoelectric layers are used, a bimorph}. As the piezoelectric material expands or contracts along its length, it forces the cantilever beam to bend, resulting in an out-of-plane deflection. Research also explores the~$d_{33}$ mode using specialized electrode configurations~\cite{155_kim_piezoelectric_2009,153_kim_effects_2015}.

{Because of the ferroelectric nature of the materials, a polarization switch occurs when the sign of the actuation voltage changes. This can be reduced by applying a direct current (DC) bias.}~\cite{6_rusconi_micro_2022}

\subsection{Electrostatic}\label{sec:electrostatic}
A third actuation principle employs electrostatic forces between charged electrodes. When a voltage is applied across electrodes separated by a dielectric medium, an electrostatic force~$F_{\mathrm{el}}$ acts on the electrodes, as described in Eq.~\eqref{eq:electrostatic_force}. In this context, {$U_{\mathrm{AC}}$}~is the signal voltage amplitude of the audio signal and {$U_{\mathrm{DC}}$}~a constant bias voltage. {Under the condition $U_{\mathrm{DC}} \gg U_{\mathrm{AC}}$, the response of the actuator to the AC signal is linearized, suppressing the frequency-doubling quadratic force component relative to the linear drive term.} The remaining terms, $C$ and~$x$, represent the capacitance and the {displacement coordinate}, respectively. The derivative of the capacitance for a parallel plate capacitor with variable electrode spacing depends on the permittivity~$\varepsilon$ of the dielectric between the electrodes, the electrode area~$A$, and their separation distance~$g$~\cite{lenk_electromechanical_2011}.
\begin{align}
	F_{\mathrm{el}} &= \frac{({U_{\mathrm{AC}}} + {U_{\mathrm{DC}}})^2}{2} \, \frac{\partial C}{\partial x} \nonumber \\[6pt]
	&= \frac{({U_{\mathrm{AC}}} + {U_{\mathrm{DC}}})^2}{2} \, \frac{\varepsilon \, A}{({g} - x)^2}\label{eq:electrostatic_force}
\end{align}

The simplest configuration, shown in Fig.~\ref{fig:pull-configuration}, involves one movable and one stationary electrode. As indicated by Eq.~\eqref{eq:electrostatic_force}, the relationship between force and voltage is nonlinear, which leads to signal distortion at high voltages. This issue is mitigated by the push-pull configuration shown in Fig.~\ref{fig:push-pull-configuration}, where a movable electrode is placed between two stationary ones, resulting {in Eq.~\eqref{eq:electrostatic_force_differential}. The force response becomes linear when {$x \ll g$}, a condition achieved through either small displacements or a large electrode gap. This simplification leads to the linear voltage-to-force relationship shown in Eq.~\eqref{eq:electrostatic_force_differential_small_displacement}.} An alternative approach uses overlapping electrodes that slide relative to each other. {In this configuration, the active capacitor area varies linearly with displacement. Because $dC/dx$ is constant, the resulting electrostatic force is independent of displacement~\cite{adams_introductory_2010}. While this displacement-independent force eliminates position-dependent nonlinearities, differential driving schemes are still required to linearize the quadratic voltage-to-force relationship.} In practice, comb structures are employed to enhance electrostatic force by increasing the effective capacitor area.

\begin{align}
	{F_{\mathrm{el}}} &{= \frac{\varepsilon \, A}{2} \, \left( \frac{(U_{\mathrm{AC}} + U_{\mathrm{DC}})^2}{(g - x)^2} - \frac{(U_{\mathrm{AC}} - U_{\mathrm{DC}})^2}{(g + x)^2} \right)}\label{eq:electrostatic_force_differential} \\[6pt]
	{F_{\mathrm{el}}} &{= \frac{2 \, \varepsilon \, A \, U_{\mathrm{AC}} \, U_{\mathrm{DC}}}{g^2}}\label{eq:electrostatic_force_differential_small_displacement}
\end{align}

{Electrostatic actuators are susceptible to pull-in instability, where the movable electrode snaps onto a stationary one, often damaging the device. This occurs when Eqs.~\eqref{eq:force_pull-in} and \eqref{eq:stiffness_pull-in} are fulfilled, where $F_{\mathrm{el}}$ is the electrostatic force as described by Eq.~\eqref{eq:electrostatic_force} and $F_\mathrm{m} = k_\mathrm{m} \, x$ the mechanical restoring force for the static case without external forces. At this point, the negative effective electrostatic stiffness~$k_{\mathrm{el}} = dF_\mathrm{el}/dx$ exceeds the mechanical restoring stiffness~$k_\mathrm{m}$ of the spring. For a parallel-plate actuator, inserting Eq.~\eqref{eq:electrostatic_force} into Eq.~\eqref{eq:force_pull-in} leads to a displacement-dependent expression for the pull-in voltage~$U_{\mathrm{PI}}$. Evaluating this expression at the pull-in displacement~$x = g/3$ leads to the pull-in voltage given in Eq.~\eqref{eq:pull-in_voltage}. The full derivation can be found in~\cite{kempe_inertial_2011} and a detailed review of the pull-in phenomenon in MEMS is provided by Zhang et al.}~\cite{zhang_electrostatic_2014}.

\begin{align}
	F_\mathrm{m} &= F_{\mathrm{el}} \label{eq:force_pull-in} \\[6pt]
	k_\mathrm{m} &= k_{\mathrm{el}} \label{eq:stiffness_pull-in} \\[6pt]
	{U_{\mathrm{PI}}} &{= \sqrt{\frac{8 \, k_{\mathrm{m}} \, g^3}{27 \, \varepsilon \, A}}} \label{eq:pull-in_voltage}
\end{align}

\section{Research Approaches}
{Having established measurement methods and the fundamental principles of sound generation and actuation, the following sections systematically survey research advances across the different categories of MEMS speakers. A summary of discussed qualitative metrics is shown in Tab.~\ref{tab:qualitative_comparison}, while all reported frequency responses are included in the supplementary material.}

As this review focuses on MEMS transducers designed for audio applications, articles are categorized based on title, abstract, and figures to assess relevance. Comparative data are drawn from research articles, conference papers, and datasheets. Conference contributions duplicating journal articles from the same research group are excluded, as are studies on piezoelectric and capacitive micromachined ultrasonic transducers, which target medical imaging and material analysis. Publications focused solely on materials, fabrication technologies, or unrelated applications such as distance sensing or medical engineering are also excluded. Devices are classified as MEMS speakers when micromachining techniques constitute a significant part of the fabrication process.

\subsection{Electrodynamic Baseband-Displacement Speakers}
{Most electrodynamic MEMS speakers follow the structure shown in Fig.~\ref{fig:schematic_electrodynamic_speaker}, typically using polymer diaphragms made of polyimide~\cite{167_shearwood_applications_1996,58_harradine_micro-machined_1997,77_cheng_silicon_2004,96_ayatollahi_materials_2009,94_majlis_compact_2017} or highly compliant PDMS to maximize displacement and SPL~\cite{88_chen_optimized_2009,87_chen_low-power_2011,95_setiarini_novel_2018}. Superior acoustic performance in terms of SPL per unit area at~\SI{1}{\kilo\hertz} (Tab.~\ref{tab:electrodynamic}) has been achieved through piston-like motion designs. These utilize either rigid, low-mass silicon diaphragms with structured backsides suspended by silicon springs~\cite{90_lemarquand_mems_2012,1_shahosseini_optimization_2013,89_sturtzer_high_2013}, silicon diaphragms with surrounding polymer suspension rings~\cite{145_neri_novel_2011}, or localized central stiffening structures~\cite{188_fang_innovative_2026}. While a magnetostrictive concept has also demonstrated high output, it requires a large external magnet, limiting practical applicability~\cite{138_s_albach_sound_2011}.}

Achieving strong magnetic fields in compact speakers requires rare earth materials, which are costly and associated with environmental and ethical concerns. {The lack of commercially widespread, standardized fabrication processes for permanent magnets restricts scalability. Consequently, external magnets are needed, increasing device size and weight while complicating integration due to the variety of required materials and components.} Coil fabrication via molding or sputtering remains challenging but has been demonstrated~\cite{1_shahosseini_optimization_2013}, while magnet assembly remains manual~\cite{86_je_compact_2009}. Compatibility with the driving circuits designed for conventional loudspeakers simplifies system integration, but the aforementioned challenges regarding manufacturability will limit scalability and automated manufacturing at an industrial scale.

Electrodynamic {MEMS speakers have demonstrated} high power density, large mechanical displacement, low actuation voltage, and quasi-linear behavior~\cite{1_shahosseini_optimization_2013,89_sturtzer_high_2013}. Although early studies identified magnetic actuation as promising, interest has declined, with two recent publications in 2026 suggesting renewed attention. As summarized in the compilation of key performance metrics in Tab.~\ref{tab:electrodynamic}, these devices achieve moderate to high SPL values, with the highest average SPL per unit area among all principles. Power consumption ranges from \SIrange{1}{500}{\milli\watt}, higher than piezoelectric devices but lower than {thermoacoustic} ones. Device areas typically fall in the range between~\SI{5}{\milli\meter\squared} and~\SI{180}{\milli\meter\squared}. They are well suited for the audible frequency range, with the fundamental resonance typically ranging from~\SIrange{0.3}{20}{\kilo\hertz}. Primary application areas include hearing aids and full-range speakers for mobile devices.

\subsection{Piezoelectric Baseband-Displacement Speakers}

\begin{figure*}[hbt!]
	\centering
	\subfloat[]{
		\begin{minipage}[b]{0.3\linewidth}
			\def\svgwidth{\columnwidth}
			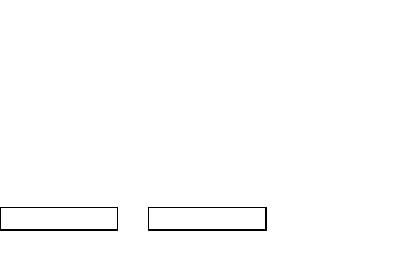
			\label{fig:schematic_electrodynamic_speaker}
		\end{minipage}
	}
	\hfill
	\subfloat[]{
		\begin{minipage}[b]{0.3\linewidth}
			\def\svgwidth{\columnwidth}
			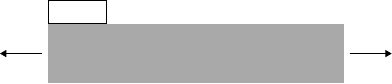
			\label{fig:piezoelectric_drive_d33}
		\end{minipage}
	}
	\hfill
	\subfloat[]{
		\begin{minipage}[b]{0.3\linewidth}
			\def\svgwidth{\columnwidth}
			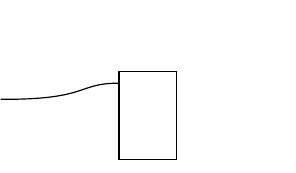
			\label{fig:70_principle}
		\end{minipage}
	}
	\vspace{0.5cm}
	\subfloat[]{
		\begin{minipage}[b]{0.3\linewidth}
			\def\svgwidth{\columnwidth}
			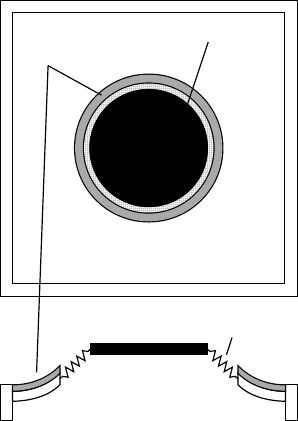
			\label{fig:schematic_piezoelectric_spring_speaker}
		\end{minipage}
	}
	\hfill
	\subfloat[]{
		\begin{minipage}[b]{0.3\linewidth}
			\def\svgwidth{\columnwidth}
			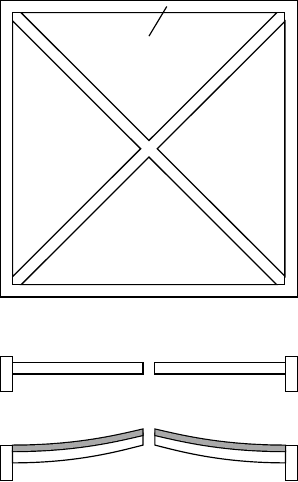
			\label{fig:schematic_piezoelectric_speaker}
		\end{minipage}
	}	
	\hfill
	\subfloat[]{
		\begin{minipage}[b]{0.3\linewidth}
			\def\svgwidth{\columnwidth}
			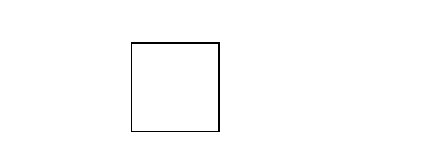
			\label{fig:peripheral_pull-configuration}
		\end{minipage}
	}
	\caption[]{\textbf{Schematic structures of MEMS speakers with different actuation concepts.}
		\subref{fig:schematic_electrodynamic_speaker}~Typical schematic structure of an electrodynamic MEMS speaker.
		\subref{fig:piezoelectric_drive_d33}~Schematic illustration of the use of the $d_{33}$ mode of a piezoelectric material for actuation through interdigitated electrodes~\cite{155_kim_piezoelectric_2009,153_kim_effects_2015}.	
		\subref{fig:70_principle}~Schematic illustration of the structure and actuation scheme from Ref.~\cite{70_liechti_piezoelectric_2022,72_liechti_high_2023,200_liechti_mems_2024}.
		\subref{fig:schematic_piezoelectric_spring_speaker}~Schematic of a piezoelectric speaker using springs and additional cantilever actuators as used in~\cite{78_cheng_design_2020,8_lin_bandwidth_2024,82_hu_design_2025,83_lin_spring_2025}.
		\subref{fig:schematic_piezoelectric_speaker}~Typical structure of a piezoelectric speaker with triangular bending cantilevers with the actuation scheme illustrated below.
		\subref{fig:peripheral_pull-configuration}~Structure of an electrostatic speaker using peripheral electrode actuation~\cite{160_sano_electret-augmented_2020,97_garud_novel_2020}.
	}
\end{figure*}

{Research on piezoelectric baseband-displacement MEMS speakers, beginning in 1996 with notable growth since 2022, is summarized in Tab.~\ref{tab:piezoelectric}. Early designs predominantly used fully clamped diaphragms~\cite{61_cheol-hyun_han_fabrication_1999,74_ko_micromachined_2003,168_cheol-hyun_han_parylene-diaphragm_2000,156_kim_improvement_2012,153_kim_effects_2015,130_zhu_study_2005,154_kim_high_2009,62_yi_micromachined_2005,65_yi_piezoelectric_2008,132_seo_micromachined_2007,63_yi_performance_2009,155_kim_piezoelectric_2009}, where the most performant devices use tensile diaphragm materials to maximize displacement and SPL per active area at~\SI{1}{\kilo\hertz}~\cite{140_cho_piezoelectrically_2010,156_kim_improvement_2012,63_yi_performance_2009,104_wang_high-spl_2020}. To enhance electromechanical coupling, some architectures enable activation of the $d_{33}$ mode from a lateral electric field by interdigitated electrodes as shown in Fig.~\ref{fig:piezoelectric_drive_d33}. This mode has a higher piezoelectric coefficient than the commonly used $d_{31}$ mode, shown in Fig.~\ref{fig:piezoelectric_drive}, enabling larger deflections at equivalent driving voltages. Only one design employing this approach achieved high performance, likely due to its larger actuator layer~\cite{153_kim_effects_2015,155_kim_piezoelectric_2009}. Among devices achieving an SPL per active area $\ge \SI{90}{\decibel}$ at~\SI{1}{\kilo\hertz}, fully clamped designs represent the second-largest structural group in the literature. However, most fully clamped devices exhibit lower performance, which explains the shift toward alternative geometries.}

{To overcome these limitations, exceptional performance ($\ge \SI{90}{\decibel}$ at~\SI{1}{\kilo\hertz}) has been achieved by separating the piezoelectric drive and diaphragm into distinct vertical layers (Fig.~\ref{fig:70_principle}). The superior performance stems from the resulting piston-like motion of the rigid diaphragm, which maximizes the effective area~\cite{70_liechti_piezoelectric_2022,144_hirano_pzt_2022,72_liechti_high_2023,200_liechti_mems_2024}. This decoupled approach is also commercially utilized by \textit{USound}~\cite{15_noauthor_usound_nodate,32_noauthor_achelous_2023-1,31_noauthor_achelous_2023,33_noauthor_adap_2023,41_noauthor_conamara_2024-2}. Similarly, for transducers with a central diaphragm supported by perimeter springs as shown in Fig.~\ref{fig:schematic_piezoelectric_spring_speaker}, piston-like displacement is achieved~\cite{64_chen_edge-released_2012,78_cheng_design_2020,163_massimino_ultrasonic_2022,8_lin_bandwidth_2024,83_lin_spring_2025}. However, performance is limited by reduced diaphragm area and leakage through inter-spring slits.}

{Furthermore, research from 2020 to 2026 is dominated by cantilever-based architectures, typically employing triangular or trapezoidal geometries similar to Fig.~\ref{fig:schematic_piezoelectric_speaker}~\cite{177_cerini_high-performance_2025,184_perli_pushpull_2026,187_gazzola_high_2026,199_zhang_piezoelectric_2026,189_deng_high-performance_2026,108_chen_design_2025,191_liu_pinwheel-shaped_2026,192_liu_bandwidth_2026,174_zheng_high-spl_2025}, although early works utilized rectangular cantilevers~\cite{166_seung_s_lee_piezoelectric_1996,60_lee_piezoelectric_1998,129_ren_tian-ling_design_2002}. Variations of the concept include different numbers of cantilevers, centrally connected elements enabling piston-like motion and requiring segmented electrodes for out-of-phase actuation~\cite{105_ma_pzt_2023,38_deng_optimization_2024}, and bimorph rather than unimorph actuators to increase displacement at the expense of fabrication complexity~\cite{114_lang_piezoelectric_2022,110_chen_monolithic_2025}. To guide design optimization of piezoelectric cantilever speakers, figures of merit were proposed, resulting in a simplified design process~\cite{176_sun_figure_2025}. Despite the efforts, only two cantilever-based designs have successfully achieved an SPL per active area $\ge \SI{90}{\decibel}$ at~\SI{1}{\kilo\hertz}~\cite{187_gazzola_high_2026,184_perli_pushpull_2026}. Commercially, the architecture is used by \textit{xMEMS}~\cite{34_noauthor_innovative_2023}, integrated into commercial products as tweeters paired with conventional dynamic drivers~\cite{43_noauthor_xmems_2024}.}

{Cantilever designs require effective leakage mitigation, as slits lead to acoustic short-circuiting between the front and back of the speaker, reducing SPL at low frequencies~\cite{165_chen_improving_2025,108_chen_design_2025,83_lin_spring_2025,182_zheng_spl_2025,191_liu_pinwheel-shaped_2026}. This issue is worsened during operation, as deflections enlarge slits, while residual stress in piezoelectric thin films such as \ce{AlN} causes static prebending~\cite{2_liu_ultrahigh-sensitivity_2022}. To address this, mitigation strategies include the mechanically open, acoustically closed (MOAC) approach~\cite{andreas_analysis_2017,56_stoppel_new_2018,andreas_design_2019,80_tseng_sound_2020,55_gazzola_design_2023,11_stoppel_highly_2025}, which relies on narrow slit dimensions to suppress airflow via viscous boundary layers. However, because this acoustic resistance is frequency-dependent, low-frequency leakage may persist if the slit width is not sufficiently small. Alternatively, sealing using flexible layers such as Parylene-C or PDMS can be employed to completely seal the device~\cite{119_wang_obtaining_2021,105_ma_pzt_2023,115_xu_piezoelectric_2023,9_wang_unsealed_2024,109_wang_capillary_2024,106_zheng_ultra-high_2025,82_hu_design_2025,174_zheng_high-spl_2025,189_deng_high-performance_2026}.}

{Multi-way operation improves SPL over a broader frequency range by combining structural elements with different resonance frequencies. It has been implemented using cantilevers of different shapes and sizes~\cite{57_stoppel_novel_2017,81_wang_multi-way_2021,7_chen_design_2023,111_fei_performance_2024,79_cheng_thd_2024,178_chen_design_2025,197_li_nested_2026}, different clamping configurations, or selectively connected groups of cantilevers~\cite{110_chen_monolithic_2025,175_lin_performance_2025,180_tsai_monolthic_2025,182_zheng_spl_2025,198_lin_bandwidth_2026}.}

{Most biomimetic designs show acoustic performance inferior to the best performing devices~\cite{183_shao_fixed-end_2026,185_shao_center-bound_2026,186_wang_flexible_2026}. A notable exception employs six triangular cantilevers with honeycomb perforations that reduce stiffness and increase PDMS deformation near the clamped region, enhancing viscoelastic damping and thereby suppressing resonance peaks and reducing THD~\cite{194_wang_biomimetic-inspired_2026}.}

{Reliability is increasingly evaluated via drop tests and long-term playback, showing no SPL degradation after impact and a reduction {of SPL} to approximately~\SI{90}{\percent} after~\SI{100}{\hour} of operation~\cite{183_shao_fixed-end_2026,185_shao_center-bound_2026,186_wang_flexible_2026,197_li_nested_2026}.}

{Piezoelectric speakers offer a middle ground in manufacturability, with materials well integrated into microfabrication and mostly CMOS compatible, though not fully manufacturable through CMOS processes~\cite{gemelli_recent_2023}. Diaphragms are typically made from silicon~(\ce{Si}), \ce{SiN}, Parylene-C, or polyimide. While early designs used \ce{ZnO} and \ce{AlN} as piezoelectric material, PZT remains widely used due to its superior piezoelectric properties, enabling high displacement at low voltages~\cite{73_kuentz_knn_2024}. It is typically deposited as sol-gel film or ceramic, with ceramics offering higher piezoelectric coefficients but requiring a non-silicon {adhesion layer}~\cite{104_wang_high-spl_2020}. The lead content of PZT and lead magnesium niobate-lead titanate (PMN-PT)~\cite{155_kim_piezoelectric_2009,153_kim_effects_2015} poses environmental and health risks~\cite{ibn-mohammed_life_2018}. Alternatives such as \ce{ZnO} and \ce{AlN} offer CMOS compatibility~\cite{113_fawzy_design_2021}, lower thermal budgets during manufacturing, and improved linearity~\cite{114_lang_piezoelectric_2022}. Doping with scandium further enhances \ce{AlN} performance~\cite{110_chen_monolithic_2025}. KNN is another lead-free option with comparable or superior properties~\cite{73_kuentz_knn_2024,148_gao_study_2014}, though its environmental footprint is higher than that of~PZT~\cite{ibn-mohammed_life_2018}. Based on the mature processes and CMOS compatibility, cost‑efficient, highly automated manufacturing at industrial scale appears feasible.}

{While the inverse piezoelectric effect allows for low-voltage operation, piezoelectric speakers typically have capacitances one order of magnitude higher than those of electrostatic speakers, requiring larger driving currents. The average SPL per unit area is slightly lower than that of electrostatic devices, but with better peak performance reaching up to~\SI{123}{\decibel} at~\SI{1}{\kilo\hertz}. Current research often revisits established concepts or focuses on aspects not captured by SPL metrics, such as sealing strategies for improving low-frequency performance or lead-free materials. Key applications include in-ear headphones and mobile phone speakers, while the absence of magnets enables additional use cases such as MRI-compatible headphone systems.}

\subsection{Electrostatic Baseband-Displacement Speakers}
{The development of electrostatic MEMS speakers began in 1996~\cite{59_rangsten_electrostatically_1996}. Early concepts often utilized pre-bent diaphragms using reduced cavity pressure~\cite{76_neumann_cmos-mems_2002,84_roberts_electrostatically_2007}. While one such design achieved one of the highest reported SPL per active area at~\SI{1}{\kilo\hertz} (Tab.~\ref{tab:electrostatic}), it required a high signal voltage amplitude $U_{\mathrm{AC}}$ of \SI{200}{\volt}~\cite{84_roberts_electrostatically_2007}. Initial attempts to repurpose electrostatic MEMS microphones for sound generation suffered from lower acoustic performance~\cite{92_glacer_reversible_2013,91_glacer_silicon_2014}, whereas the transition to structures optimized for acoustic actuation improved performance.}

{To overcome the limitations of conventional parallel-plate actuators, peripheral electrode actuation as shown in Fig.~\ref{fig:peripheral_pull-configuration} reduces squeeze-film damping while increasing pull-in voltage and maximum deflection compared to full counter electrodes. This approach has enabled one of the highest SPL per active area at~\SI{1}{\kilo\hertz} reported for electrostatic MEMS speakers~\cite{160_sano_electret-augmented_2020,97_garud_novel_2020}. It can be combined with an electret layer to lower the required bias voltage by generating a built-in electric field, functioning as the electrostatic equivalent of permanent magnets~\cite{160_sano_electret-augmented_2020}. Comparable acoustic performance has been achieved with a unique passive concept that integrates a cantilever array inside the slit of an electrostatic slot dipole or PCB antenna~\cite{181_ruiz_integrating_2025,190_ruiz_towards_2026}. Here, an amplitude-modulated radio signal directly drives the cantilevers via electrostatic forces, generating sound without requiring an external power source.}

{Furthermore, designs with push-pull actuation have demonstrated high acoustic output~\cite{85_hanseup_kim_bi-directional_2005} while enhancing sound quality by reducing THD~\cite{85_hanseup_kim_bi-directional_2005,91_glacer_silicon_2014,69_kaiser_push-pull_2022}. Alternatively, vertical parallel-plate capacitors arranged along the diaphragm edges can be used to induce out-of-plane motion~\cite{35_bhuiyan_electrostatic_2024}. In contrast to these out-of-plane architectures, in-plane displacement designs utilize vertically oriented, moving cantilevers instead of a single diaphragm~\cite{68_kaiser_concept_2019,69_kaiser_push-pull_2022}. While increasing the cantilever height enhances displaced volume and SPL without enlarging chip area, the approach remains bound by micro-fabrication constraints, and published performance remained below that of out-of-plane designs.}

{An analysis of devices with the highest SPL per active area at~\SI{1}{\kilo\hertz} reveals that they are structurally diverse, making it difficult to isolate common structural drivers responsible for their exceptional acoustic output. Consequently, a wide variety of diaphragm materials is also used, including gold~\cite{142_murarka_printed_2016}, \ce{SiC}~\cite{84_roberts_electrostatically_2007}, polymers such as Parylene-C~\cite{85_hanseup_kim_bi-directional_2005} and graphene monolayers on polymers~\cite{158_khan_mutually_2020}.} Most electrostatic MEMS speakers benefit from CMOS compatibility~\cite{59_rangsten_electrostatically_1996,68_kaiser_concept_2019,69_kaiser_push-pull_2022}, enabling automated mass-manufacturing and scalability. Exceptions include the use of polymers for diaphragms~\cite{76_neumann_cmos-mems_2002} or acoustic sealing. 

Anodic corrosion under humidity and electric fields is a known reliability concern~\cite{18_shea_effects_2004,22_liu_corrosion_2011,30_stolz_reliability_2021}. Although this phenomenon has been extensively studied, it is rarely addressed in the MEMS speaker literature. Mitigation strategies include passivation coatings and corrosion-sensitive warning structures~\cite{18_shea_effects_2004,22_liu_corrosion_2011,30_stolz_reliability_2021}.

As compiled in the performance overview in Tab.~\ref{tab:electrostatic}, these devices achieve a higher mean SPL and SPL per active area than piezoelectric and {thermoacoustic} alternatives. Their resonance frequencies are also the highest among all principles. Electrostatic speakers typically exhibit low capacitance in the range of~\si{\pico\farad}, resulting in a lower current {draw} compared to piezoelectric devices. Target applications are primarily in-ear headphones.

\subsection{{Ultrasound-Based Speakers}}
{Ultrasound-based speakers using digital sound reconstruction (DSR) were introduced in 2002 and use arrays of electrostatic, out-of-plane actuated diaphragms to achieve 8-bit signal reconstruction~\cite{17_diamond_digital_2003}. The precision of micromachining helps minimize variations between active speaklets and thereby reduces distortion. Besides electrostatic actuation~\cite{171_de_pasquale_modeling_2016}, actuator arrays for DSR were also realized using PZT as piezoelectric material~\cite{25_dejaeger_development_2012,49_casset_256_2015,28_29_arevalo_mems_2016}. However, the SPL per active area at \SI{1}{\kilo\hertz} (Tab.~\ref{tab:ultrasound}) of standard DSR remains below that of high-performing baseband-displacement speakers. }

{Improvements to DSR include a single diaphragm with segmented electrodes for individually addressable out-of-plane actuation~\cite{170_santos_microelectromechanical_2014} and non-uniform pulses to reduce the number of required speaklets for a given bitrate~\cite{172_monkronthong_multiple-level_2014,173_monkronthong_study_2016}. Another enhancement in the form of advanced digital sound reconstruction (ADSR) employs speaklets with side channels and shutters, allowing the diaphragm to return to its rest position without generating reverse airflow~\cite{4_mayrhofer_new_2021}. The principle was validated using a macroscopic pump and remains at the proof-of-concept stage.}

{An alternative ultrasound-based approach is the carrier-based speaker developed by \textit{SonicEdge}~\cite{10_chen_modulated_2025}, comprising an array of electrostatically driven pumps. Each pump comprises two movable diaphragms and a fixed diaphragm between them, with a slit at its center. The bottom diaphragm follows the carrier frequency of~\SI{400}{\kilo\hertz}, amplitude modulated with the audio signal, while the top diaphragm demodulates by modulating the acoustic impedance between the middle cavity and ear volume. The SPL per active area at~\SI{1}{\kilo\hertz} of this device is in the range of high-performing baseband-displacement variants, with particularly superior performance at low frequencies. In contrast, a similar concept employing twelve independent units consisting of four piezoelectric \ce{AlN} triangular cantilever actuators and a stacked dual-cantilever demodulator demonstrated significantly lower acoustic performance~\cite{195_zheng_ultrasonic_2025}.}

Additional research opportunities arise regarding ultrasound exposure, although no health risks are reported at hundreds of kilohertz~\cite{10_chen_modulated_2025}. Existing studies and regulatory limits primarily address occupational ultrasonic noise~\cite{smagowska_effects_2013}, indicating both auditory and non-auditory effects, including thermal effects, subjective symptoms, and functional changes. In the mid-frequency range of \SIrange{50}{500}{\kilo\hertz}, subjective symptoms are not expected, making this range suitable for speakers. At SPLs above \SI{150}{\decibel}, thermal and mechanical effects can occur. Furthermore, the acoustic impedance mismatch between air and human tissue limits ultrasonic energy penetration, which is advantageous for in-ear applications. Ideally, the {safety} of {ultrasound-based} speakers should be confirmed with dedicated clinical studies, especially considering ultrasound emission directly in the ear canal at frequencies relevant for pulse-based speakers. As explained in Sec.~\ref{sec:ultrasound}, the ultrasonic energy content depends on the modulation scheme, motivating the exploration of alternative modulation techniques from signal theory.

{To expand bandwidth, integrating ultrasound-based woofers with baseband-displacement tweeters on the same chip represents a promising research direction. However, co-fabricating both transducer topologies increases process complexity and may necessitate performance trade-offs. Furthermore, the high operating frequencies of ultrasound-based speakers in general demand frequent electrical charging cycles, leading to increased power consumption compared to baseband concepts. While this limitation can be partially mitigated by charge-recuperation driving circuits, the overall efficiency remains bound to the carrier frequency and drive principle.}

{Despite these efficiency trade-offs, ultrasound-based speakers are especially suited for low-frequency sound generation. Because of the inverse frequency dependence of the acoustic output, more ultrasound pulses fit into one period of the audio signal, enabling SPLs approaching \SI{140}{\decibel} at low frequencies, which are required for active noise cancellation~\cite{visse_microphones_2025}. While the currently scarce literature prevents a comprehensive comparison with other sound generation concepts, the available performance metrics are compiled in Tab.~\ref{tab:ultrasound}.}

\subsection{{Thermoacoustic Speakers}}
Early {thermoacoustic} speakers employed silicon, either as porous tracks~\cite{123_shinoda_thermally_1999} or as a nanocrystalline layer on a monocrystalline substrate~\cite{122_kiuchi_new_2005}. A key development was the introduction of a DC bias superimposed on the AC audio signal to suppress frequency doubling~\cite{122_kiuchi_new_2005}.

{Maximizing SPL per active area at~\SI{1}{\kilo\hertz} (Tab.~\ref{tab:thermoacoustic}) requires transduction materials with a low heat capacity per unit area, making carbon nanotubes and graphene particularly suitable. Graphene is the most commonly used material for thermoacoustic speakers, typically deposited as single- or multi-layer films on various substrates~\cite{23_tian_graphene--paper_2011,98_tian_single-layer_2012,118_suk_thermoacoustic_2012}. However, configurations with substrates generally exhibit lower acoustic performance due to thermal conduction losses, with polyimide substrates being a notable exception~\cite{159_wang_research_2020}.}

{To minimize these substrate-induced thermal losses, the highest efficiencies are achieved using either highly porous or entirely substrate-free approaches. Implementations include woven graphene fabrics suspended over porous substrates~\cite{99_zhang_high-performance_2017} and self-supporting, multi-layered carbon nanotube films~\cite{137_xiao_flexible_2008}. Alternatively, suspended metal wire arrays offer comparable acoustic performance while maintaining full compatibility with standard cleanroom processes~\cite{141_niskanen_suspended_2009,vesterinen_fundamental_2010}.}

Fabrication methods vary from simple processes~\cite{123_shinoda_thermally_1999,122_kiuchi_new_2005,141_niskanen_suspended_2009} to complex transfer techniques~\cite{23_tian_graphene--paper_2011,24_tian_flexible_2011,118_suk_thermoacoustic_2012,100_fei_lowvoltage_2014,159_wang_research_2020}, depending on the materials involved. This limits conclusions on suitability for highly automated industrial-scale manufacturing. The materials enable flexible, transparent, or large-area devices suitable for integration into clothing and other non-planar surfaces~\cite{23_tian_graphene--paper_2011,98_tian_single-layer_2012,118_suk_thermoacoustic_2012,137_xiao_flexible_2008}. Research activity peaked between 2008 and 2012, with no new publications since 2020.

Regarding performance, {thermoacoustic} speakers present a unique trade-off between structural flexibility and power efficiency. The resulting performance metrics, which are compiled in Tab.~\ref{tab:thermoacoustic}, illustrate that while high absolute SPLs can be achieved, overall efficiency remains low due to large device areas and the absence of resonant amplification. Drive voltages are relatively low, but power consumption is significantly higher than in other actuation principles.

\section{Performance Analysis across the MEMS Speaker Research Landscape}
To synthesize the diverse research approaches on MEMS speakers, this section evaluates chronological trends and frequency response profiles. As a structural foundation for this analysis, Tab.~\ref{tab:principles_vs_drive_concepts_without_thermoacoustic} shows how research activity is distributed across different drive principles and sound generation concepts, highlighting the dominant approaches.

\subsection{SPL and Normalized Pressure over Years of Publication}
Fig.~\ref{fig:spl_over_years} presents the evolution of published absolute sound pressure levels (SPL) at \SI{1}{\kilo\hertz} over the years of publication. Values are extracted from publications where speakers were built, measured, and measurement methods reported. To enable comparison across concepts, values are converted to ear coupler equivalents with Eq.~\eqref{eq:equivalent_SPL}, under the assumptions discussed in Sec.~\ref{sec:acoustic_parameters}. Despite increasing publication activity, the average SPL has remained largely constant. {However, when focusing on the best-performing devices, the stagnation in maximum achievable SPL is less apparent.} Piezoelectric speakers initially showed rising SPL values until around 2017, followed by stagnation. This trend could be reflected by a shift in research priorities toward miniaturization, reduced input voltage, and mitigation of low-frequency leakage. {These are factors} not captured by a single SPL value. Electrostatic speakers exhibited early growth, then declined due to impractical initial designs, but have recently gained attention through novel concepts such as {ultrasound-based} sound generation. Electrodynamic and {thermoacoustic} speakers show limited recent activity, likely due to poor compatibility with MEMS fabrication. While {thermoacoustic} devices can achieve high SPLs, this typically requires large form factors. Tab.~\ref{tab:principles_vs_drive_concepts_without_thermoacoustic} confirms the dominance of piezoelectric actuators with {baseband displacement}, which account for more publications than all other principles combined. This is presumably due to the ability to reach high actuation forces at moderate voltages. {Ultrasound-based} speakers remain underrepresented due to their recent emergence.

Where available, measured SPL values are converted to pressure amplitude~$\hat{p}$ and normalized by the active area~$A_\mathrm{eff}$ and {peak-to-peak} signal voltage amplitude~$U_{\mathrm{AC}}$. The resulting graph in Fig.~\ref{fig:normalized_pressure_active_area_over_years} reveals a general increase in normalized sound pressure over the years of publication, possibly reflecting ongoing miniaturization trends. Piezoelectric speakers show mixed performance in recent years, with a few high-performing designs but many near the average, suggesting that only specific implementations yield significant benefits, or their advancements are not captured by this metric of comparison. Electrostatic speakers initially performed above average, then declined, and have recently improved in normalized pressure, largely due to {ultrasound-based} concepts using electrostatic actuation. While some piezoelectric devices still achieve higher normalized pressure, electrostatic approaches may surpass them with further development. It is important to note that this comparison is limited to a single frequency and does not consider the entire frequency response.

\subsection{SPL and THD Response Profiles}
The extracted SPL response profiles are averaged over octave bands for seven time intervals indicated by color and marker style. For datasets with more than two curves, standard deviation is shown by horizontal bars. Separate diagrams are used for ear coupler and free-field measurements, as no established method currently exists for converting free-field SPL to equivalent SPL in an ear coupler over the entire frequency range. SPL measured with an ear coupler is scaled to a standard volume of \SI{1.26}{\cubic\cm} using Eq.~\eqref{eq:conversion_coupler_volumes}, while free-field measurements are scaled to a reference distance of \SI{1}{\cm} using Eq.~\eqref{eq:distance_free-field}, assuming far-field sound propagation.

\begin{figure*}[hbt!]
	\centering
	\subfloat[]{
		\begin{minipage}[b]{0.45\linewidth}
			\includegraphics[width=1\columnwidth]{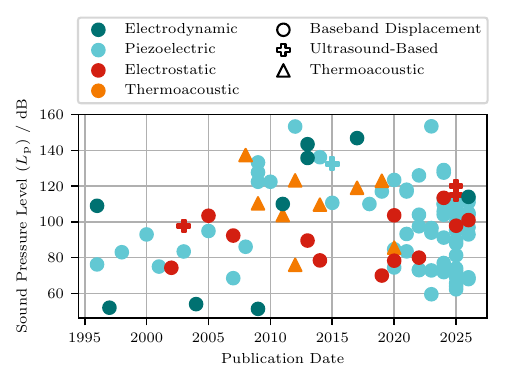}
			\label{fig:spl_over_years}
			\vspace*{-1.5em}
		\end{minipage}
	}
	\hfill
	\subfloat[]{
		\begin{minipage}[b]{0.45\linewidth}
			\includegraphics[width=1\columnwidth]{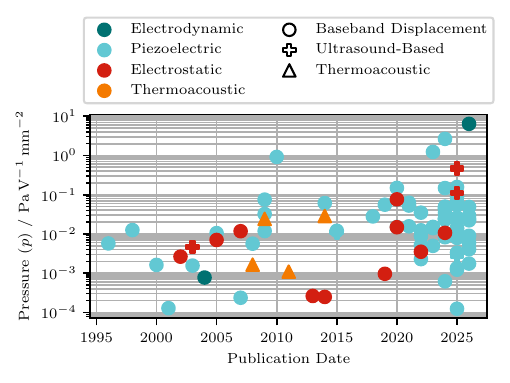}
			\label{fig:normalized_pressure_active_area_over_years}
			\vspace*{-1.5em}
		\end{minipage}
	}
	\vspace{-0.3cm}
	\subfloat[]{
		\begin{minipage}[b]{0.45\linewidth}
			\includegraphics[width=\columnwidth]{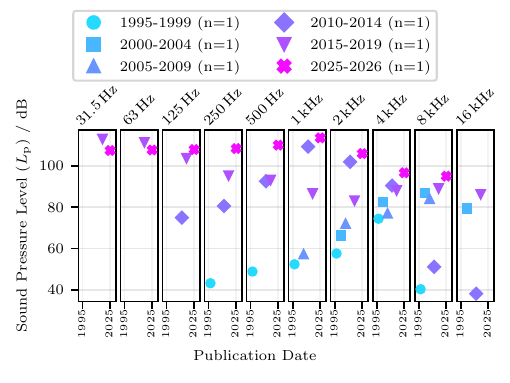}
			\label{fig:spl_behavior-Electrodynamic_ear_coupler}
			\vspace*{-1.5em}
		\end{minipage}
	}
	\hfill
	\subfloat[]{
		\begin{minipage}[b]{0.45\linewidth}
			\includegraphics[width=1\columnwidth]{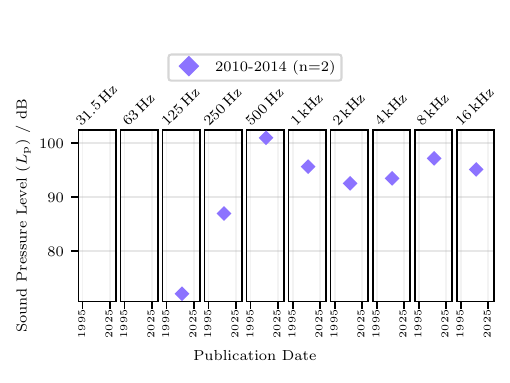}
			\label{fig:spl_behavior-Electrodynamic_free_field}
			\vspace*{-1.5em}
		\end{minipage}
	}
	\vspace{-0.3cm}
	\subfloat[]{
		\begin{minipage}[b]{0.45\linewidth}
			\includegraphics[width=1\columnwidth]{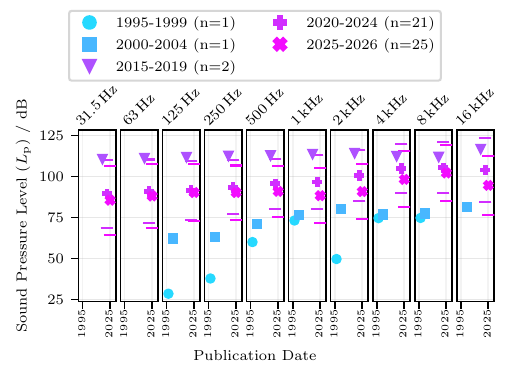}
			\label{fig:spl_behavior-Piezoelectric_ear_coupler}
			\vspace*{-1.5em}
		\end{minipage}
	}
	\hfill
	\subfloat[]{
		\begin{minipage}[b]{0.45\linewidth}
			\includegraphics[width=1\columnwidth]{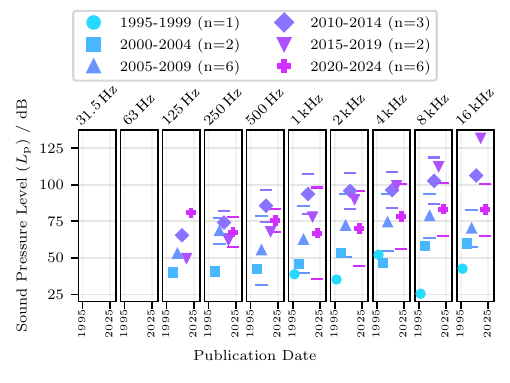}
			\label{fig:spl_behavior-Piezoelectric_free_field}
			\vspace*{-1.5em}
		\end{minipage}
	}
	\caption[]{\textbf{SPL and normalized pressure over the year of publication and averaged SPL response behavior grouped by drive principle and by measurement setup.}
		\subref{fig:spl_over_years}~SPL at a frequency of \SI{1}{\kilo\hertz} as a function of publication year with color indicating the drive principle and marker shape indicating the sound generation concept.
		\subref{fig:normalized_pressure_active_area_over_years}~Normalized pressure per active area and {peak-to-peak} signal voltage amplitude at a frequency of \SI{1}{\kilo\hertz} as a function of publication year with color indicating the drive principle and marker shape indicating the sound generation concept.
		\subref{fig:spl_behavior-Electrodynamic_ear_coupler}~{Averaged SPL per octave band for the specified time intervals for published electrodynamic MEMS speakers measured in an ear coupler, normalized to \SI{1.26}{\cubic\cm} with Eq.~\eqref{eq:conversion_coupler_volumes}. Raw data are included in the supplementary material.}
		\subref{fig:spl_behavior-Electrodynamic_free_field}~{Averaged SPL per octave band for the specified time intervals for published electrodynamic MEMS speakers measured under free-field conditions, normalized to \SI{1}{\cm} distance according to Eq.~\eqref{eq:distance_free-field}. Raw data are included in the supplementary material.}
		\subref{fig:spl_behavior-Piezoelectric_ear_coupler}~{Averaged SPL per octave band for the specified time intervals for published piezoelectric MEMS speakers measured in an ear coupler, normalized to \SI{1.26}{\cubic\cm} with Eq.~\eqref{eq:conversion_coupler_volumes}. Raw data are included in the supplementary material.}
		\subref{fig:spl_behavior-Piezoelectric_free_field}~{Averaged SPL per octave band for the specified time intervals for published piezoelectric MEMS speakers measured under free-field conditions, normalized to \SI{1}{\cm} distance according to Eq.~\eqref{eq:distance_free-field}. Raw data are included in the supplementary material.}}
\end{figure*}

The averaged SPL of electrodynamic MEMS speakers measured in an ear coupler (Fig.~\ref{fig:spl_behavior-Electrodynamic_ear_coupler}) indicates a performance increase over time, although the dataset is limited. The frequency response varies strongly between devices. Reported SPL values span~\SIrange{40}{115}{\decibel}, largely depending on resonance frequency, although limited sample sizes prevent robust conclusions. In free-field conditions (Fig.~\ref{fig:spl_behavior-Electrodynamic_free_field}), available data are even more scarce and indicate poor low-frequency performance.

For piezoelectric MEMS speakers measured in an ear coupler (Fig.~\ref{fig:spl_behavior-Piezoelectric_ear_coupler}), the frequency response improved from 1995 to 2019, followed by a slight degradation. This trend is consistent with a shift in research focus away from absolute performance. Responses remain relatively flat, but recent devices show increased variability, as indicated by a high standard deviation. Below~\SI{100}{\hertz}, a moderate SPL reduction suggests leakage effects. Reported SPL values span~\SIrange{60}{120}{\decibel}, with the lowest value corresponding to the first device reported in 1996~\cite{166_seung_s_lee_piezoelectric_1996}, which was not optimized for sound radiation. In free-field measurements (Fig.~\ref{fig:spl_behavior-Piezoelectric_free_field}), performance improved until 2019 and subsequently decreased, while maintaining improved low-frequency behavior but exhibiting high variability.

\begin{figure*}[hbt!]
	\centering
	\subfloat[]{
		\begin{minipage}[b]{0.45\linewidth}
			\includegraphics[width=1\columnwidth]{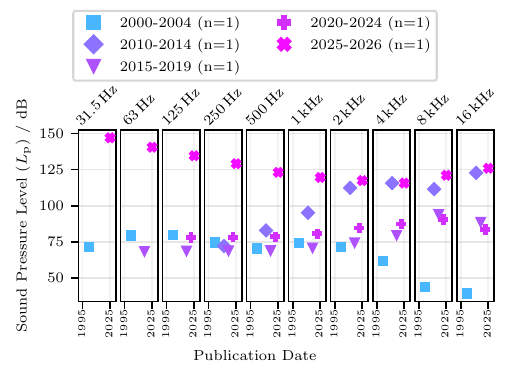}
			\label{fig:spl_behavior-Electrostatic_ear_coupler}
			\vspace*{-1.5em}
		\end{minipage}
	}
	\hfill
	\subfloat[]{
		\begin{minipage}[b]{0.45\linewidth}
			\includegraphics[width=1\columnwidth]{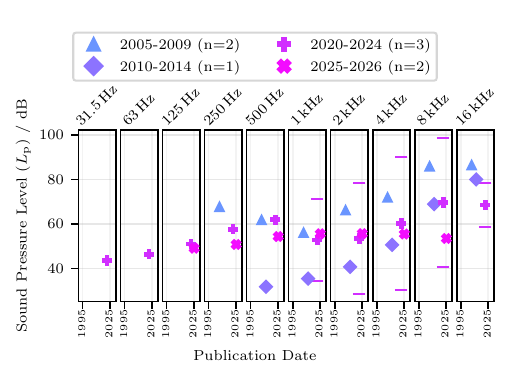}
			\label{fig:spl_behavior-Electrostatic_free_field}
			\vspace*{-1.5em}
		\end{minipage}
	}
	\vspace{0.5cm}
	\subfloat[]{
		\begin{minipage}[b]{0.45\linewidth}
			\includegraphics[width=1\columnwidth]{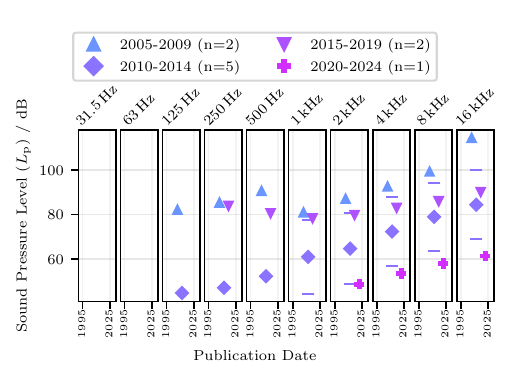}
			\label{fig:spl_behavior-Thermoacoustic_free_field}
			\vspace*{-1.5em}
		\end{minipage}
	}
	\hfill
	\subfloat[]{
		\begin{minipage}[b]{0.45\linewidth}
			\includegraphics[width=1\columnwidth]{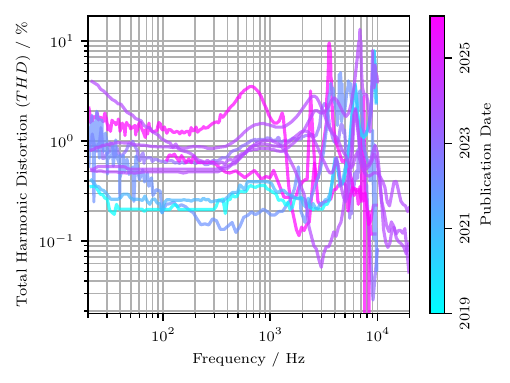}
			\label{fig:drive_concept_colored_thd_curves}
			\vspace*{-1.5em}
		\end{minipage}
	}
	\caption[]{\textbf{Averaged SPL response behavior and THD response profiles from MEMS speaker literature.}
		\subref{fig:spl_behavior-Electrostatic_ear_coupler}~{Averaged SPL per octave band for the specified time intervals for published electrostatic MEMS speakers measured in an ear coupler, normalized to \SI{1.26}{\cubic\cm} with Eq.~\eqref{eq:conversion_coupler_volumes}. Raw data are included in the supplementary material.}
		\subref{fig:spl_behavior-Electrostatic_free_field}~{Averaged SPL per octave band for the specified time intervals for published electrostatic MEMS speakers measured under free-field conditions, normalized to \SI{1}{\cm} distance according to Eq.~\eqref{eq:distance_free-field}. Raw data are included in the supplementary material.}
		\subref{fig:spl_behavior-Thermoacoustic_free_field}~{Averaged SPL per octave band for the specified time intervals for published {thermoacoustic} speakers measured under free-field conditions, normalized to \SI{1}{\cm} distance according to Eq.~\eqref{eq:distance_free-field}. Raw data are included in the supplementary material.}
		\subref{fig:drive_concept_colored_thd_curves}~THD response profiles at a reference SPL of \SI{94}{\decibel} over frequency colored by {year of publication}. Raw data are included in the supplementary material.}
\end{figure*}

The frequency responses of electrostatic MEMS speakers measured in an ear coupler (Fig.~\ref{fig:spl_behavior-Electrostatic_ear_coupler}) indicate a general performance improvement over time, although the dataset is very limited. A notable outlier is the only {ultrasound-based} device for which data are available~\cite{10_chen_modulated_2025}. This device represents the most recent time interval and exhibits exceptional low-frequency performance. Reported SPL values range from~\SI{20}{\decibel} at~\SI{20}{\kilo\hertz} for an early MEMS design~\cite{76_neumann_cmos-mems_2002} to~\SI{150}{\decibel} at~\SI{20}{\hertz} for the {ultrasound-based} approach. In free-field conditions (Fig.~\ref{fig:spl_behavior-Electrostatic_free_field}), performance varies across time intervals and shows high variability.

As shown in the average ear-coupler SPL responses for electrodynamic (Fig.~\ref{fig:spl_behavior-Electrodynamic_ear_coupler}), piezoelectric (Fig.~\ref{fig:spl_behavior-Piezoelectric_ear_coupler}), and electrostatic transducers (Fig.~\ref{fig:spl_behavior-Electrostatic_ear_coupler}), no {baseband-displacement} speaker achieves \SI{120}{\decibel} at low frequencies, and performance further degrades in the presence of leakage. The only exception is the {ultrasound-based} device reported in~\cite{10_chen_modulated_2025}, indicated by the pink markers in Fig.~\ref{fig:spl_behavior-Electrostatic_ear_coupler}.

For {thermoacoustic} speakers, only free-field data are available (Fig.~\ref{fig:spl_behavior-Thermoacoustic_free_field}), and no clear trend is observed over the publication dates.

Regarding free-field measurements in general, it can be noted that few speakers were measured below~\SI{100}{\hertz}. This is consistent with the theoretical behavior of a mechanical resonator below its fundamental resonance, where the acceleration, and consequently the SPL, is expected to decrease with a slope of \SI{40}{\decibel} per decade {with decreasing frequencies}. As expected, SPLs recorded under free-field conditions are generally lower than those measured in ear couplers.

Nonlinearities are assessed using total harmonic distortion (THD), as described in Sec.~\ref{sec:acoustic_parameters}, with values for different speakers at \SI{1}{\kilo\hertz} in Tab.~\ref{tab:thd}. It should be noted that the THD values offer limited comparability, as it is common to compare THD at distinct SPLs, such as \SI{94}{\decibel}, while such data are sparsely represented in the literature. Furthermore, distortions can have large variations over the frequency spectrum, while Tab.~\ref{tab:thd} contains single frequency data. Therefore, the frequency-dependent behavior shown in Fig.~\ref{fig:drive_concept_colored_thd_curves}, measured at~\SI{94}{\decibel} SPL at~\SI{1}{\kilo\hertz}, can provide more insights. Because all devices measured under these conditions are piezoelectric speakers, the data indicate generally lower distortion below~\SI{400}{\hertz}, with only two exceptions. At higher frequencies, distortion increases, with THD peaks occurring {at subharmonics}~\cite{2_liu_ultrahigh-sensitivity_2022, 11_stoppel_highly_2025, 184_perli_pushpull_2026}.

\section{Outlook \& Conclusion}
The emergence of MEMS speakers is driven by advantages in semiconductor manufacturing, including precise tolerances, scalability, and integration with drive electronics. This review provides a structured comparison of MEMS speaker technologies based on the sound generation concepts of {baseband displacement}, {ultrasound-based} sound generation, and {thermoacoustics}, along with the drive principles such as electrodynamic, piezoelectric, and electrostatic.

Across the different concepts, progress has been made in increasing sound pressure normalized to active area and drive voltage, reducing THD, and advancing miniaturization. However, meeting the SPL requirements of applications like TWS across the full audible range remains challenging, particularly at low frequencies under real-world conditions, such as when an earbud has an imperfect seal within the user's ear. {Ultrasound-based} speakers, {particularly carrier-based concepts}, offer a promising solution as they can achieve high SPLs at low frequencies. However, these concepts are still emerging, and many design aspects remain unexplored, including optimal speaker geometry and the choice between piezoelectric and electrostatic actuation. Additionally, systematic studies on the effects of ultrasound and novel applications present research opportunities. It is also unclear whether these speakers can function as full-range devices or if combining them with {baseband-displacement} units could improve the frequency response. Their performance under free-field conditions requires further evaluation.

For electrodynamic MEMS speakers, further research could be conducted on manufacturing methods for permanent magnets. However, electrodynamic speakers continue to suffer from assembly challenges, limiting scalability despite their attractive acoustic performance. {Thermoacoustic} speakers, typically fabricated over large areas, present an opportunity for adaptation to in-ear formats. However, they exhibit poor low-frequency performance and high power consumption. Their use may therefore remain limited to applications requiring transparency or mechanical flexibility. While piezoelectric MEMS speakers have been extensively studied, identifying lead-free materials with properties comparable to PZT offers potential for further investigations. Promising candidates include \ce{AlScN} and KNN. For electrostatic MEMS speakers, reducing the required drive voltage, {effectively improving sensitivity,} remains an area open for further research. Additionally, environmental robustness could be investigated, prioritizing mitigation strategies that avoid substances of concern for environmental and health safety. Research could focus on durability and on the effects of environmental exposure on SPL, THD, and long-term performance. Currently, no standardized tests exist to assess environmental robustness specifically for MEMS speakers, highlighting the need for dedicated reliability protocols, including accelerated aging tests. Mitigation approaches include passivation strategies and packaging. However, packaging is inherently constrained because MEMS speakers must remain acoustically exposed during operation. As packaging materials influence both biocompatibility and acoustic performance, their effects should be systematically investigated in future studies. Because no dedicated listener studies have explicitly examined MEMS speakers, future studies could evaluate the perceptual performance of these devices.

{Beyond the individual drive principles, a broader research opportunity exists in exploring combinations of sound generation concepts and actuation mechanisms that remain largely unexplored. To date, published ultrasound-based MEMS speakers have been realized using piezoelectric and electrostatic actuation, while reported electrodynamic and thermoacoustic implementations focus on baseband sound generation. Extending digital sound reconstruction or carrier-based concepts to thermoacoustic and electrodynamic transducers could reveal alternative performance trade-offs and further broaden the design space of MEMS speakers.}

Despite substantial progress, many opportunities remain in this interdisciplinary field, bridging materials science, microfabrication, acoustics, and circuit design. MEMS speakers have the potential to transform consumer electronics, medical devices, and industrial systems. This review provides researchers and industry experts with a global overview of the state of the art. This synthesis highlights progress in miniaturization, performance, and integration, while illuminating persistent challenges and promising opportunities. Ultimately, collaborative efforts across these disciplines will be crucial to realizing the full potential of MEMS speakers for a new generation of compact, high-fidelity audio devices.

\bibliographystyle{naturemag}
\bibliography{apssamp}

\begin{acknowledgments}
The IPCEI ME/CT project is supported by the Federal Ministry for Economic Affairs and Climate Action on the basis of a decision by the German Parliament, by the Ministry for Economic Affairs, Labor and Tourism of Baden-Württemberg based on a decision of the State Parliament of Baden-Württemberg, the Free State of Saxony on the basis of the budget adopted by the Saxon State Parliament, the Bavarian State Ministry for Economic Affairs, Regional Development and Energy and financed by the European Union -- NextGenerationEU.
\end{acknowledgments}

\section*{Competing Interests}
Nils Wittek, Anton Melnikov, and Bert Kaiser are employed by Bosch Sensortec GmbH. The company had no influence on the content or conclusions of this review. The other authors declare no competing interests.

\nolinenumbers

\begin{table*}[hbt!]
	\caption{\label{tab:qualitative_comparison}{Qualitative comparison of MEMS speaker concepts and their characteristics.}}
	\begin{ruledtabular}
		\begin{tabular}{c|c|c|c|c|c}
			& \multicolumn{3}{c|}{{Baseband Displacement}} & & \\
			\makecell{Comparison\\Indicators} & Electrodynamic & Piezoelectric & Electrostatic & {Thermoacoustic} & \makecell{{Ultrasound-}\\{Based}} \\ \hline
			Advantages & \makecell[l]{\textbullet~Similar to\\~~\,conventional\\~~\,designs} & \makecell[l]{\textbullet~Low voltages \\ \textbullet~Improvements via \\~~\,bimorphs or $d_{33}$\\~~\,actuation} & \makecell[l]{\textbullet~Low capacitances} & \makecell[l]{\textbullet~Structural\\~~\,flexibility\\\textbullet~Optical\\~~\,transparency} & \makecell[l]{\textbullet~Different concepts \\ \textbullet~Best low-frequency\\~~\,performance} \\ \hline
			Disadvantages & \makecell[l]{\textbullet~Require rare\\~~\,earth materials\\ \textbullet~High power\\~~\,consumption} & \makecell[l]{\textbullet~Most used material\\~~\,contains lead} & \makecell[l]{\textbullet~Lower force output \\ \textbullet~High voltages \\ \textbullet~Environmental\\~~\,robustness concerns} & \makecell[l]{\textbullet~Low efficiency} & \makecell[l]{\textbullet~Ultrasound energy\\~~\,content \\ \textbullet~Not easy to simulate} \\ \hline
			\makecell{Target\\Applications} & \makecell[l]{\textbullet~Hearing aids\\ \textbullet~Mobile phone\\~~\,speakers} & \makecell[l]{\textbullet~In-ear headphones\\ \textbullet~Mobile phone\\~~\,speakers} & \makecell[l]{\textbullet~In-ear headphones} & \makecell[l]{\textbullet~Special applications} & \makecell[l]{\textbullet~In-ear headphones}
		\end{tabular}
	\end{ruledtabular}
\end{table*}

\begin{table*}[hbt!]
	\caption{\label{tab:electrodynamic}Comparison of electrodynamic {baseband-displacement} MEMS speakers at \SI{1}{\kilo\hertz}, measured in free-field (ff) closed chamber (cc), {or according to IEC 60318-4}. $L_{\mathrm{p}{\mathrm{ec}}}$ denotes the equivalent SPL in a standard ear coupler, calculated using Eq.~\eqref{eq:equivalent_SPL} for free-field data and Eq.~\eqref{eq:conversion_coupler_volumes} for other measurement volumes. {$L_{\mathrm{p}_{\mathrm{eff}}}$ is the SPL normalized to an active area~$A_{\mathrm{eff}}$ of \SI{1}{\square\milli\meter}}, $A_{\mathrm{total}}$ the total device area, and $f_{\mathrm{res}}$ the first resonance frequency. Rows are sorted in descending order by {$L_{\mathrm{p}_{\mathrm{eff}}}$}, where available.}
	\begin{ruledtabular}
		\begin{tabular}{ccccccccc}
			Article & $L_{\mathrm{p}}$ / $\si{\decibel}$ & Measurement &  $L_{\mathrm{p}_{\mathrm{ec}}}$ / $\si{\decibel}$ & $\bm{L_{\mathrm{p}_{\mathrm{eff}}}}$ / $\si{\decibel}$ & $A_{\mathrm{eff}}$ / $\si{\mm\squared}$ & $A_{\mathrm{total}}$ / $\si{\mm\squared}$ & Bandwidth / $\si{\kilo\hertz}$ & $f_{\mathrm{res}}$ / $\si{\hertz}$ \\ \hline
			Shahosseini~\cite{1_shahosseini_optimization_2013} & 80 & ff (\SI{10}{\cm}) & 143 & 99 & 177 & - & \numrange{0.33}{70} & 480 \\
			{Fang}~\cite{188_fang_innovative_2026} & 114 & IEC 60318-4 & 114 & 94 & 9.62 & - & \numrange{0.02}{10} & - \\
			Sturtzer~\cite{89_sturtzer_high_2013} & 92 & ff (\SI{1}{\cm}) & 136 & 91 & 177 & - & \numrange{0.1}{20} & 335 \\
			Chen~\cite{87_chen_low-power_2011} & 106 & cc (\SI{2}{\cubic\cm}) & 110 & 90 & 10 & - & \numrange{0.125}{16} & 1360 \\
			Albach~\cite{138_s_albach_sound_2011} & 95 & cc (\SI{2}{\cubic\cm}) & 99 & 82 & 7.45 & - & \numrange{0.1}{3} & 450 \\
			Shearwood~\cite{167_shearwood_applications_1996} & 105 & cc (\SI{2}{\cubic\cm}) & 109 & 75 & 50 & - & - & - \\
			Harradine~\cite{58_harradine_micro-machined_1997} & 48 & cc (\SI{2}{\cubic\cm}) & 52 & 38 & 5 & - & \numrange{0.2}{10} & 3000 \\
			Chen~\cite{88_chen_optimized_2009} & 47 & cc (\SI{2}{\cubic\cm}) & 51 & 37 & 5 & - & \numrange{1}{6} & - \\
			Cheng~\cite{77_cheng_silicon_2004} & 50 & cc (\SI{2}{\cubic\cm}) & 54 & 32 & 12 & 25 & \numrange{1}{20} & 2800 \\
			Majlis~\cite{94_majlis_compact_2017} & 85 & cc (\SI{1.5}{\cubic\cm}) & 147 & - & - & 25 & \numrange{0.02}{20} & -
		\end{tabular}
	\end{ruledtabular}
\end{table*}
\clearpage
\begin{longtable*}[hbt!]{cccccccccccc}
			\caption{\label{tab:piezoelectric}Comparison of piezoelectric {baseband-displacement} MEMS speakers at \SI{1}{\kilo\hertz}, measured in free-field (ff), closed chamber (cc), or according to IEC~60318-4. Parameters are defined in Tab.~\ref{tab:electrodynamic}, with additional values for {peak-to-peak} signal voltage amplitude {$U_{\mathrm{AC}}$}, DC bias voltage {$U_{\mathrm{DC}}$}, and capacitance $C$. Rows are sorted in descending order by {$L_{\mathrm{p}_{\mathrm{eff}}}$}, where available.}\\
			\toprule\rule{0pt}{12pt}
			Article & $L_{\mathrm{p}}$/$\si{\decibel}$ & Measurement & $L_{\mathrm{p}_{\mathrm{ec}}}$/$\si{\decibel}$ & $\bm{L_{\mathrm{p}_{\mathrm{eff}}}}$/\si{\decibel} & $A_{\mathrm{eff}}$/$\si{\mm\squared}$ & $A_{\mathrm{total}}$/$\si{\mm\squared}$ & \makecell{Band-\\width/$\si{\kilo\hertz}$} & $f_{\mathrm{res}}$/$\si{\hertz}$ & {$U_{\mathrm{AC}}$}/$\si{\volt}$ & {$U_{\mathrm{DC}}$}/$\si{\volt}$ & $C$/$\si{\nano\farad}$ \\
			\colrule
			\endfirsthead			
			\caption[]{Comparison of piezoelectric {baseband-displacement} MEMS speakers at a frequency of \SI{1}{\kilo\hertz} measured in free-field~(ff), a closed chamber~(cc), or according to IEC~60318-4 (continued)}\\
			\colrule
			Article & $L_{\mathrm{p}}$/$\si{\decibel}$ & Measurement & $L_{\mathrm{p}_{\mathrm{ec}}}$/$\si{\decibel}$ & $\bm{L_{\mathrm{p}_{\mathrm{eff}}}}$/$\si{\decibel}$ & $A_{\mathrm{eff}}$/$\si{\mm\squared}$ & $A_{\mathrm{total}}$/$\si{\mm\squared}$ & \makecell{Band-\\width/$\si{\kilo\hertz}$} & $f_{\mathrm{res}}$/$\si{\hertz}$ & {$U_{\mathrm{AC}}$}/$\si{\volt}$ & {$U_{\mathrm{DC}}$}/$\si{\volt}$ & $C$/$\si{\nano\farad}$ \\
			\colrule
			\endhead		
			\colrule
			\endfoot
			\colrule
			\botrule
			\endlastfoot
			Liechti~\cite{72_liechti_high_2023} & 110 & ff (\SI{1}{\cm}) & 153 & 122 & 36 & 100 & \numrange{0.3}{20} & 1000 & 30 & 15 & - \\
			Cho~\cite{140_cho_piezoelectrically_2010} & 79 & ff (\SI{1}{\cm}) & 122 & 113 & 3.14 & - & \numrange{0.2}{10} & - & 13 & - & - \\
			Kim~\cite{156_kim_improvement_2012} & 90 & ff (\SI{10}{\cm}) & 153 & 102 & 360 & - & \numrange{0.1}{10} & 800 & - & - & - \\
			{Liechti}~\cite{200_liechti_mems_2024} & 129 & IEC 60318-4 & 129 & 98 & 36 & - & \numrange{0.03}{20} & 1360 & 15 & 15 & - \\
			{USound} \cite{32_noauthor_achelous_2023-1} & 118 & IEC 60318-4 & 118 & 96 & 12 & 31.49 & \numrange{0.02}{20} & 2700 & 30 & 15 & 27 \\
			Yi~\cite{63_yi_performance_2009} & 79 & ff (\SI{1}{\cm}) & 122 & 95 & 25 & - & \numrange{0.1}{15} & - & 20 & - & - \\
			{USound} \cite{31_noauthor_achelous_2023} & 117 & IEC 60318-4 & 117 & 95 & 12 & 31.49 & \numrange{0.02}{20} & 2700 & 30 & 15 & 27 \\
			{USound} \cite{33_noauthor_adap_2023} & 117 & IEC 60318-4 & 117 & 95 & 12 & 31.49 & \numrange{0.02}{20} & 2900 & 30 & 15 & 26 \\
			Wang~\cite{104_wang_high-spl_2020} & 80 & ff (\SI{1}{\cm}) & 123 & 94 & 28.27 & 44.89 & \numrange{0.4}{10} & 4200 & 10 & - & - \\
			{Gazzola}~\cite{187_gazzola_high_2026} & 111 & IEC 60318-4 & 111 & 94 & - & 20.25 & \numrange{0.01}{10} & 10700 & 30 & 15 & 35 \\
			Kim~\cite{155_kim_piezoelectric_2009} & 90 & ff (\SI{1}{\cm}) & 133 & 90 & 144 & 225 & \numrange{0.1}{10} & 840 & 28 & - & 29 \\
			{USound} \cite{41_noauthor_conamara_2024-2} & 113 & IEC 60318-4 & 113 & 90 & 13.6 & 28.27 & \numrange{0.02}{40} & 2300 & 27 & 10 & 11 \\
			{Perli}~\cite{184_perli_pushpull_2026} & 113 & IEC 60318-4 & 113 & 90 & 14.44 & 20.25 & \numrange{0.1}{10} & 7500 & 30 & 15 & 50 \\
			Zheng~\cite{174_zheng_high-spl_2025} & 104 & IEC 60318-4 & 104 & 88 & 6.25 & - & \numrange{0.02}{20} & 5000 & 60 & - & - \\
			Han~\cite{143_han_thin-film_2022} & 53 & ff (\SI{30}{\cm}) & 126 & 88 & 81 & 100 & \numrange{0.1}{100} & - & 20 & - & - \\
			{USound} \cite{39_noauthor_conamara_2024-1} & 106 & IEC 60318-4 & 106 & 87 & 9.1 & 19.63 & \numrange{2}{80} & 5200 & 27 & 10 & 9 \\
			{USound} \cite{40_noauthor_conamara_2024} & 110 & IEC 60318-4 & 110 & 87 & 13.6 & 28.27 & \numrange{2}{80} & 3900 & 27 & 10 & 9 \\
			Stoppel~\cite{56_stoppel_new_2018} & 110 & IEC 60318-4 & 110 & 86 & 16 & - & \numrange{0.02}{20} & 8300 & 20 & - & - \\
			{Wang}~\cite{194_wang_biomimetic-inspired_2026} & 105 & IEC 60318-4 & 105 & 85 & 10.39 & - & \numrange{0.01}{20} & - & 20 & 6 & - \\
			Chen~\cite{165_chen_improving_2025} & 104 & IEC 60318-4 & 104 & 84 & 10.24 & - & \numrange{0.02}{20} & 4050 & 20 & - & - \\
			Wang~\cite{9_wang_unsealed_2024} & 104 & IEC 60318-4 & 104 & 84 & 10.4 & - & \numrange{0.02}{20} & 9500 & 10 & - & - \\
			Kim~\cite{154_kim_high_2009} & 84 & ff (\SI{1}{\cm}) & 128 & 83 & 182 & 440 & \numrange{0.1}{10} & 563 & 32 & - & - \\
			Zheng~\cite{106_zheng_ultra-high_2025} & 95 & IEC 60318-4 & 95 & 81 & 4.9 & 9 & \numrange{0.02}{20} & 5600 & 100 & - & 0.78 \\
			Stoppel~\cite{11_stoppel_highly_2025} & 100 & IEC 60318-4 & 100 & 81 & 9 & - & \numrange{0.02}{5} & 1500 & 2 & - & 84 \\
			Liu~\cite{2_liu_ultrahigh-sensitivity_2022} & 104 & IEC 60318-4 & 104 & 80 & 16 & - & \numrange{0.02}{20} & 3600 & 22.63 & - & - \\
			{Deng}~\cite{189_deng_high-performance_2026} & 97 & IEC 60318-4 & 97 & 78 & 9 & - & \numrange{0.02}{20} & - & 10 & 5 & - \\
			{Liu}~\cite{191_liu_pinwheel-shaped_2026} & 104 & IEC 60318-4 & 104 & 76 & 25 & - & \numrange{0.02}{20} & 8200 & 30 & 15 & - \\
			Kim~\cite{153_kim_effects_2015} & 67 & ff (\SI{1}{\cm}) & 111 & 76 & 56.75 & 136.89 & \numrange{0.1}{10} & 1840 & 14.14 & - & - \\
			Cheng~\cite{78_cheng_design_2020} & 75 & IEC 60318-4 & 75 & 75 & 1 & - & \numrange{0.02}{20} & 1850 & 2 & - & - \\
			Deng~\cite{38_deng_optimization_2024} & 91 & IEC 60318-4 & 91 & 75 & 6.25 & - & \numrange{0.02}{20} & 5600 & 20 & 2.5 & - \\
			Hirano~\cite{144_hirano_pzt_2022} & 98 & IEC 60318-4 & 98 & 73 & 16 & 49 & \numrange{0.02}{20} & 4200 & 15 & 7.5 & - \\
			Wang~\cite{81_wang_multi-way_2021} & 84 & IEC 60318-4 & 84 & 71 & 4 & - & \numrange{0.1}{10} & 1540 & 2 & - & - \\
			Lee~\cite{60_lee_piezoelectric_1998} & 46 & ff (\SI{0.5}{\cm}) & 83 & 71 & 4 & - & \numrange{0.0001}{40} & 14500 & 8 & - & - \\
			Gazzola~\cite{55_gazzola_design_2023} & 94 & IEC 60318-4 & 94 & 71 & 14.44 & - & \numrange{0.1}{10} & 10600 & 7 & - & - \\
			Lang~\cite{114_lang_piezoelectric_2022} & 73 & IEC 60318-4 & 73 & 67 & 1.96 & 16 & \numrange{0.1}{20} & 10000 & 28.28 & - & - \\
			Yi~\cite{62_yi_micromachined_2005} & 62 & ff (\SI{0.3}{\cm}) & 95 & 67 & 25 & - & \numrange{0.4}{12} & - & 6 & - & - \\
			Cheng~\cite{79_cheng_thd_2024} & 72 & IEC 60318-4 & 72 & 65 & 2.25 & 9 & \numrange{1}{8.9} & 6350 & 1 & 2 & - \\
			Lin~\cite{8_lin_bandwidth_2024} & 77 & IEC 60318-4 & 77 & 65 & 4 & 7.4 & \numrange{1}{15.4} & 6800 & 2 & 9 & - \\
			Han~\cite{168_cheol-hyun_han_parylene-diaphragm_2000} & 64 & ff (\SI{0.2}{\cm}) & 93 & 65 & 25 & - & \numrange{0.1}{20} & 1820 & 31 & - & - \\
			Liechti~\cite{136_liechti_polymer-based_2025} & 106 & IEC 60318-4 & 106 & 65 & 113.1 & 225 & \numrange{0.02}{20} & 3043 & 6 & 15 & - \\
			Lee~\cite{166_seung_s_lee_piezoelectric_1996} & 72 & cc (\SI{2}{\cubic\cm}) & 76 & 64 & 4 & 4.08 & \numrange{0.1}{20} & 890 & 8 & - & - \\
			Ko~\cite{74_ko_micromachined_2003} & 40 & ff (\SI{1}{\cm}) & 83 & 64 & 9 & - & \numrange{0}{20} & 7300 & 30 & - & - \\
			Shih~\cite{164_shih_woofer-type_2025} & 108 & IEC 60318-4 & 108 & 63 & 180 & - & \numrange{0.05}{1} & 895.4 & 30 & - & - \\
			{Zhang}~\cite{199_zhang_piezoelectric_2026} & 89 & cc (\SI{2}{\cubic\centi\meter}) & 93 & 62 & 36 & - & \numrange{0.1}{20} & 1300 & 20 & - & - \\
			Tseng~\cite{80_tseng_sound_2020} & 85 & IEC 60318-4 & 85 & 61 & 16.2 & - & \numrange{0.1}{20} & - & 2 & - & - \\
			Zheng~\cite{182_zheng_spl_2025} & 90 & IEC 60318-4 & 90 & 61 & 28.3 & - & \numrange{0.4}{20} & 3200 & 3 & - & - \\
			Yi~\cite{65_yi_piezoelectric_2008} & 53 & ff (\SI{0.3}{\cm}) & 86 & 60 & 20.25 & - & \numrange{0.4}{12} & - & 5 & - & - \\
			{Lin}~\cite{198_lin_bandwidth_2026} & 69 & IEC 60318-4 & 69 & 59 & 3.3 & - & \numrange{0.16}{22.3} & 9700 & 5.93 & 5 & - \\
			Chen~\cite{110_chen_monolithic_2025} & 74 & IEC 60318-4 & 74 & 59 & 5.85 & - & \numrange{0.02}{20} & 6820 & 20 & - & - \\
			Wang~\cite{109_wang_capillary_2024} & 74 & IEC 60318-4 & 74 & 59 & 5.85 & - & \numrange{0.02}{20} & 18200 & 40 & - & 0.5 \\
			Hu~\cite{82_hu_design_2025} & 69 & IEC 60318-4 & 69 & 57 & 4 & 9 & \numrange{0.1}{13.3} & 6500 & 2 & 10 & - \\
			Ma~\cite{105_ma_pzt_2023} & 73 & IEC 60318-4 & 73 & 57 & 6.25 & - & \numrange{0.02}{20} & 6300 & 4 & - & - \\
			{Liu}~\cite{192_liu_bandwidth_2026} & 68 & IEC 60318-4 & 68 & 56 & 4 & - & \numrange{1}{20} & 6700 & 1.98 & - & - \\
			Lin~\cite{83_lin_spring_2025} & 66 & IEC 60318-4 & 66 & 55 & 3.4 & - & \numrange{1}{20} & 12800 & 2 & 9 & - \\
			Guilvaiee~\cite{161_guilvaiee_validated_2023} & 14 & ff (\SI{1.3}{\cm}) & 60 & 51 & 2.82 & - & \numrange{1}{20} & 4000 & - & - & - \\
			Dan~\cite{75_cheol-hyun_dan_fabrication_2001} & 71 & cc (\SI{2}{\cubic\cm}) & 75 & 47 & 25 & - & \numrange{0.1}{20} & - & 50 & - & - \\
			Seo~\cite{132_seo_micromachined_2007} & 36 & ff (\SI{0.3}{\cm}) & 67 & 44 & 16 & - & \numrange{0.3}{12} & - & 20 & - & - \\
			Becker~\cite{117_becker_meander-shaped_2025} & 64 & IEC 60318-4 & 64 & 43 & 11.34 & - & \numrange{0.02}{20} & 40000 & 32 & - & - \\
			Gao~\cite{148_gao_study_2014} & 83 & ff (\SI{3.16}{\cm}) & 136 & - & - & 621 & \numrange{0.1}{20} & 700 & 14.14 & - & 194 \\
			Fei~\cite{111_fei_performance_2024} & 84 & ff (\SI{1}{\cm}) & 128 & - & - & 25 & \numrange{0.2}{20} & 1000 & 3 & - & - \\
			Cerini~\cite{177_cerini_high-performance_2025} & 112 & IEC 60318-4 & 112 & - & - & 20.25 & \numrange{0.1}{10} & 11600 & 30 & 15 & - \\
			{xMEMS} \cite{37_clara_datasheet_2024-1} & 111 & IEC 60318-4 & 111 & - & - & 16 & \numrange{0.02}{20} & 12200 & 30 & 10 & 24 \\
			{xMEMS} \cite{36_clara_datasheet_2024} & 110 & IEC 60318-4 & 110 & - & - & 19.2 & \numrange{0.02}{20} & 14700 & 30 & 10 & 42 \\
			Xu~\cite{115_xu_piezoelectric_2023} & 97 & IEC 60318-4 & 97 & - & - & - & \numrange{0.1}{10} & 5000 & 28.28 & - & - \\
			Sun~\cite{176_sun_figure_2025} & 95 & IEC 60318-4 & 95 & - & - & 7.7 & \numrange{0.1}{20} & 4500 & 7.07 & - & - \\
			Wang~\cite{179_wang_micropatterned_2025} & 81 & IEC 60318-4 & 81 & - & - & - & \numrange{0.02}{20} & - & 2 & 1 & - \\
			Pavageau~\cite{71_pavageau_highly_2022} & 30 & ff (\SI{1}{\cm}) & 73 & - & - & 4.91 & \numrange{1}{100} & 50000 & 15 & 15 & - \\
			Chen~\cite{108_chen_design_2025} & 73 & IEC 60318-4 & 73 & - & - & 12.5 & \numrange{0.1}{10} & 4400 & 2 & - & - \\
			Tsai~\cite{180_tsai_monolthic_2025} & 70 & IEC 60318-4 & 70 & - & - & 2 & \numrange{1}{20} & 8300 & 2 & 6 & - \\
			Chen~\cite{178_chen_design_2025} & 62 & IEC 60318-4 & 62 & - & - & 4 & \numrange{1}{17.8} & 8600 & 2 & 9 & - \\
			Wang~\cite{120_wang_piezoelectric_2023} & 97 & IEC 60318-4 & - & - & 12.57 & - & \numrange{0.2}{20} & - & 2 & - & - \\
			Wang~\cite{119_wang_obtaining_2021} & 93 & IEC 60318-4 & - & - & 10.39 & - & \numrange{0.02}{20} & 6700 & 2 & - & - \\
			{Li}~\cite{197_li_nested_2026} & 86 & cc & - & - & 16 & - & \numrange{0.02}{20} & - & 22.63 & - & - \\
			{Shao}~\cite{185_shao_center-bound_2026} & 80 & cc & - & - & 8 & 42 & \numrange{0.02}{20} & 5000 & 22.63 & - & - \\
			{Shao}~\cite{183_shao_fixed-end_2026} & 78 & cc & - & - & 16 & - & \numrange{0.02}{20} & 10600 & 22.63 & - & - \\
			{Wang}~\cite{186_wang_flexible_2026} & 77 & cc & - & - & 16 & - & \numrange{0.02}{20} & - & 22.63 & - & - \\
\end{longtable*}

\begin{table*}[hbt!]
	\caption{\label{tab:electrostatic}Comparison of electrostatic {baseband-displacement} MEMS speakers at \SI{1}{\kilo\hertz}, measured in free-field~(ff), a closed chamber~(cc), or according to IEC~60318-4. The listed parameters are explained in Tab.~\ref{tab:piezoelectric}. Rows are sorted in descending order by {$L_{\mathrm{p}_{\mathrm{eff}}}$}, where available.}
	\begin{ruledtabular}
		\begin{tabular}{cccccccccccc}
			Article & $L_{\mathrm{p}}$/$\si{\decibel}$ & Measurement & $L_{\mathrm{p}_{\mathrm{ec}}}$/$\si{\decibel}$ & $\bm{L_{\mathrm{p}_{\mathrm{eff}}}}$/\si{\decibel} & $A_{\mathrm{eff}}$/$\si{\mm\squared}$ & $A_{\mathrm{total}}$/$\si{\mm\squared}$ & \makecell{Band-\\width/$\si{\kilo\hertz}$} & $f_{\mathrm{res}}$/$\si{\hertz}$ & {$U_{\mathrm{AC}}$}/$\si{\volt}$ & {$U_{\mathrm{DC}}$}/$\si{\volt}$ & $C$/$\si{\nano\farad}$ \\ \hline
			Garud~\cite{97_garud_novel_2020} & 60 & ff (\SI{1}{\cm}) & 104 & 98 & 1.89 & - & \numrange{0.2}{20} & - & 30 & 30 & - \\		
			Roberts~\cite{84_roberts_electrostatically_2007} & 49 & ff (\SI{1}{\cm}) & 92 & 98 & 0.5 & 7.5 & \numrange{1}{20} & 54500 & 200 & - & - \\
			Ruiz~\cite{181_ruiz_integrating_2025} & 54 & ff (\SI{1}{\cm}) & 98 & 95 & 1.44 & - & \numrange{0.1}{10} & 1400 & - & - & - \\
			Kim~\cite{85_hanseup_kim_bi-directional_2005} & 60 & ff (\SI{1}{\cm}) & 103 & 91 & 4 & 15 & \numrange{1}{25} & - & 150 & - & - \\
			Bhuiyan~\cite{35_bhuiyan_electrostatic_2024} & 70 & ff (\SI{1}{\cm}) & 113 & 85 & 25 & - & \numrange{0.02}{20} & 7900 & 50 & - & - \\
			Sano~\cite{160_sano_electret-augmented_2020} & 31 & ff (\SI{1.5}{\cm}) & 78 & 68 & 3.14 & - & \numrange{1}{100} & - & 5 & 40 & - \\
			Neumann~\cite{76_neumann_cmos-mems_2002} & 76 & cc (\SI{1}{\cubic\cm}) & 74 & 68 & 1.96 & - & \numrange{0.02}{10} & - & 28.6 & 67 & - \\
			Kaiser~\cite{69_kaiser_push-pull_2022} & 80 & IEC 60318-4 & 80 & 64 & 19.8 & 110 & \numrange{0.02}{20} & 4200 & 12 & 20 & 0.26 \\
			Kaiser~\cite{68_kaiser_concept_2019} & 70 & IEC 60318-4 & 70 & 51 & 9.3 & - & \numrange{0.05}{20} & 9100 & 10 & 40 & 0.065 \\
			Glacer~\cite{92_glacer_reversible_2013} & 86 & cc (\SI{2}{\cubic\mm}) & 90 & 45 & 162.22 & 460.8 & \numrange{0.2}{20} & - & 20 & 20 & - \\
			Glacer~\cite{91_glacer_silicon_2014} & 15 & ff (\SI{10}{\cm}) & 78 & 42 & 63.71 & 364.29 & \numrange{0.4}{20} & - & 15 & 10 & 0.015 \\
			{Ruiz}~\cite{190_ruiz_towards_2026} & 58 & ff (\SI{1}{\cm}) & 101 & - & - & - & \numrange{0.1}{10} & 1200 & - & - & -
		\end{tabular}
	\end{ruledtabular}
\end{table*}

\begin{table*}[hbt!]
	\caption{\label{tab:ultrasound}Comparison of {ultrasound-based} MEMS speakers at \SI{1}{\kilo\hertz}, measured in free-field~(ff) or according to IEC~60318-4. The listed parameters are explained in Tab.~\ref{tab:piezoelectric}. Rows are sorted in descending order by {$L_{\mathrm{p}_{\mathrm{eff}}}$}, where available.}
	\begin{ruledtabular}
		\begin{tabular}{cccccccccccc}
			Article & $L_{\mathrm{p}}$/$\si{\decibel}$ & \makecell{Measure-\\ment} & $L_{\mathrm{p}_{\mathrm{ec}}}$/$\si{\decibel}$ & $\bm{L_{\mathrm{p}_{\mathrm{eff}}}}$/\si{\decibel} & $A_{\mathrm{eff}}$/$\si{\mm\squared}$ & $A_{\mathrm{total}}$/$\si{\mm\squared}$ & \makecell{Band-\\width/$\si{\kilo\hertz}$} & $f_{\mathrm{res}}$/$\si{\hertz}$ & {$U_{\mathrm{AC}}$}/$\si{\volt}$ & {$U_{\mathrm{DC}}$}/$\si{\volt} $ & \makecell{Concept\\Type} \\ \hline
			Chen~\cite{10_chen_modulated_2025} & 120 & \makecell{IEC\\60318-4} & 120 & 100 & 10.34 & 11.2 & \numrange{0.01}{20} & 400000 & 25 & 25 & Pump \\
			Diamond~\cite{17_diamond_digital_2003} & 97.8 & \makecell{IEC\\60318-4} & 97.8 & 76 & 11.9 & 27 & \numrange{0.05}{4} & - & 40 & - & DSR \\
			{Zheng}~\cite{195_zheng_ultrasonic_2025} & 106 & \makecell{IEC\\60318-4} & 106 & 75 & 36 & - & \numrange{0.02}{10} & 47000 & 56.57 & - & Pump \\
			Casset~\cite{49_casset_256_2015} & 66.81 & ff (\SI{13}{\cm}) & 132.54 & 70 & 1359.18 & 1600 & \numrange{0.5}{50} & 17300 & 8 & - & DSR \\
			\makecell{{SonicEdge}\\SC200\cite{45_46_47_noauthor_se_nodate}} & 123 & \makecell{IEC\\60318-4} & 123 & - & - & 49 & \numrange{0.02}{50} & - & - & - & Pump \\
			\makecell{{SonicEdge}\\ST100\cite{45_46_47_noauthor_se_nodate}} & 115 & \makecell{IEC\\60318-4} & 115 & - & - & 28 & \numrange{0.005}{50} & - & - & - & Pump \\
			\makecell{{SonicEdge}\\SC100\cite{45_46_47_noauthor_se_nodate}} & 115 & \makecell{IEC\\60318-4} & 115 & - & - & 26 & \numrange{0.02}{50} & - & - & - & Pump
		\end{tabular}
	\end{ruledtabular}
\end{table*}

\begin{table*}[hbt!]
	\caption{\label{tab:thermoacoustic}Comparison of {thermoacoustic} speakers at \SI{1}{\kilo\hertz} measured in free-field~(ff). The listed parameters are explained in Tab.~\ref{tab:piezoelectric}, with addition of the electrical power $P_{\mathrm{el}}$. Rows are sorted in descending order by {$L_{\mathrm{p}_{\mathrm{eff}}}$}, where available.}
	\begin{ruledtabular}
		\begin{tabular}{ccccccccc}
			Article & $L_{\mathrm{p}}$ / $\si{\decibel}$ & Measurement & $L_{\mathrm{p}_{\mathrm{ec}}}$ / $\si{\decibel}$ & $\bm{L_{\mathrm{p}_{\mathrm{eff}}}}$/\si{\decibel} & $A_{\mathrm{eff}}$ / $\si{\mm\squared}$ & $A_{\mathrm{total}}$ / $\si{\mm\squared}$ & Bandwidth / $\si{\kilo\hertz}$ & $P_{\mathrm{el}}$ / $\si{\watt}$ \\ \hline
			Zhang~\cite{99_zhang_high-performance_2017} & 76 & ff (\SI{1}{\cm}) & 119 & 79 & 100 & - & \numrange{}{20} & 1 \\
			Xiao~\cite{137_xiao_flexible_2008} & 80 & ff (\SI{5}{\cm}) & 137 & 78 & 900 & - & \numrange{0.01}{30} & 12 \\
			Niskanen~\cite{141_niskanen_suspended_2009} & 50 & ff (\SI{7}{\cm}) & 110 & 76 & 50 & - & \numrange{0.6}{20} & 17 \\
			Wang~\cite{101_wang_study_2019} & 50 & ff (\SI{30}{\cm}) & 123 & 75 & 250 & - & \numrange{0.001}{200} & - \\
			Fei~\cite{100_fei_lowvoltage_2014} & 57 & ff (\SI{3}{\cm}) & 110 & 70 & 100 & - & \numrange{1}{20} & 0.1 \\
			Tian~\cite{98_tian_single-layer_2012} & 66 & ff (\SI{5}{\cm}) & 123 & 55 & 2581 & - & \numrange{20}{50} & 1 \\
			Tian~\cite{24_tian_flexible_2011} & 46 & ff (\SI{5}{\cm}) & 104 & 44 & {1000} & - & \numrange{0.1}{50} & - \\
			Suk~\cite{118_suk_thermoacoustic_2012} & 23 & ff (\SI{3}{\cm}) & 76 & 36 & 100 & - & \numrange{1}{20} & 0.25 \\
			Wang~\cite{159_wang_research_2020} & 33 & ff (\SI{3}{\cm}) & 86 & 27 & 900 & - & - & -
		\end{tabular}
	\end{ruledtabular}
\end{table*}

\begin{table*}[hbt!]
	\caption{\label{tab:principles_vs_drive_concepts_without_thermoacoustic}Number of displacement-based sound generation concepts and drive principles.}
	\begin{ruledtabular}
		\begin{tabular}{cccc}
			\diagbox{Sound Generation Principle}{Drive Concept} & Electrodynamic & Electrostatic &  Piezoelectric \\ \hline
			{Baseband Displacement} & 15 & 19 & 89 \\
			{Ultrasound-Based} & 0 & 11 & 7
		\end{tabular}
	\end{ruledtabular}
\end{table*}

\begin{table*}[hbt!]
	\caption{\label{tab:thd}Comparison of THD values at a frequency of \SI{1}{\kilo\hertz} measured in free-field~(ff), or according to IEC~60318-4. $L_{\mathrm{p}_{\mathrm{ec}}}$ is the equivalent SPL in a standard ear coupler calculated with Eq.~\eqref{eq:equivalent_SPL} from free-field measurement and Eq.~\eqref{eq:conversion_coupler_volumes} from another measurement volume. Rows are sorted in ascending order by THD.}
	\begin{ruledtabular}
		\begin{tabular}{ccccccc}
			Source & $L_{\mathrm{p}_{\mathrm{ec}}}$ / $\si{\decibel}$ & Measurement & \textbf{THD} / $\si{\percent}$ & Drive Principle & Sound Generation Concept & \makecell{SPL of THD\\measurement / $\si{\decibel}$} \\ \hline
			Sturtzer~\cite{89_sturtzer_high_2013} & 136 & ff (\SI{1}{\cm}) & 0.014 & Electrodynamic & {Baseband Displacement} & - \\
			Stoppel~\cite{11_stoppel_highly_2025} & 100 & IEC 60318-4 & 0.2 & Piezoelectric & {Baseband Displacement} & 94 \\
			{USound} \cite{31_noauthor_achelous_2023} & 117 & IEC 60318-4 & 0.3 & Piezoelectric & {Baseband Displacement} & 94 \\
			Becker~\cite{117_becker_meander-shaped_2025} & 64 & IEC 60318-4 & 0.3 & Piezoelectric & {Baseband Displacement} & - \\
			Xiao~\cite{137_xiao_flexible_2008} & 137 & ff (\SI{5}{\cm}) & 0.31 & Thermoacoustic & {Thermoacoustic} & - \\
			Zheng~\cite{182_zheng_spl_2025} & 90 & IEC 60318-4 & 0.36 & Piezoelectric & {Baseband Displacement} & - \\
			{USound} \cite{32_noauthor_achelous_2023-1} & 118 & IEC 60318-4 & 0.4 & Piezoelectric & {Baseband Displacement} & 94 \\
			Cerini~\cite{177_cerini_high-performance_2025} & 112 & IEC 60318-4 & 0.46 & Piezoelectric & {Baseband Displacement} & 94 \\
			{xMEMS} \cite{37_clara_datasheet_2024-1} & 111 & IEC 60318-4 & 0.5 & Piezoelectric & {Baseband Displacement} & 94 \\
			Liu~\cite{2_liu_ultrahigh-sensitivity_2022} & 104 & IEC 60318-4 & 0.5 & Piezoelectric & {Baseband Displacement} & 94 \\
			Chen~\cite{165_chen_improving_2025} & 104 & IEC 60318-4 & 0.52 & Piezoelectric & {Baseband Displacement} & 108 \\
			Gazzola~\cite{55_gazzola_design_2023} & 94 & IEC 60318-4 & 0.61 & Piezoelectric & {Baseband Displacement} & - \\
			Wang~\cite{109_wang_capillary_2024} & 74 & IEC 60318-4 & 0.728 & Piezoelectric & {Baseband Displacement} & - \\
			Chen~\cite{110_chen_monolithic_2025} & 74 & IEC 60318-4 & 0.77 & Piezoelectric & {Baseband Displacement} & - \\
			{USound} \cite{40_noauthor_conamara_2024} & 110 & IEC 60318-4 & 0.9 & Piezoelectric & {Baseband Displacement} & 94 \\
			Lang~\cite{114_lang_piezoelectric_2022} & 73 & IEC 60318-4 & 0.9 & Piezoelectric & {Baseband Displacement} & - \\
			{USound} \cite{41_noauthor_conamara_2024-2} & 113 & IEC 60318-4 & 1 & Piezoelectric & {Baseband Displacement} & 94 \\
			{xMEMS} \cite{36_clara_datasheet_2024} & 110 & IEC 60318-4 & 1 & Piezoelectric & {Baseband Displacement} & 94 \\
			Stoppel~\cite{56_stoppel_new_2018} & 110 & IEC 60318-4 & 1 & Piezoelectric & {Baseband Displacement} & 85 \\
			Diamond~\cite{17_diamond_digital_2003} & 98 & IEC 60318-4 & 1 & Electrostatic & {Ultrasound-Based} & - \\			
			Xu~\cite{115_xu_piezoelectric_2023} & 97 & IEC 60318-4 & 1.07 &  Piezoelectric & {Baseband Displacement} & 94 \\
			Hu~\cite{82_hu_design_2025} & 69 & IEC 60318-4 & 1.28 & Piezoelectric & {Baseband Displacement} & - \\
			{USound} \cite{33_noauthor_adap_2023} & 117 & IEC 60318-4 & 1.4 & Piezoelectric & {Baseband Displacement} & 94 \\
			{USound} \cite{39_noauthor_conamara_2024-1} & 106 & IEC 60318-4 & 1.4 & Piezoelectric & {Baseband Displacement} & 94 \\
			Cheng~\cite{79_cheng_thd_2024} & 72 & IEC 60318-4 & 1.57 & Piezoelectric & {Baseband Displacement} & - \\
			Lin~\cite{83_lin_spring_2025} & 66 & IEC 60318-4 & 1.65 & Piezoelectric & {Baseband Displacement} & - \\
			Deng~\cite{38_deng_optimization_2024} & 91 & IEC 60318-4 & 1.86 & Piezoelectric & {Baseband Displacement} & - \\
			Kaiser~\cite{69_kaiser_push-pull_2022} & 80 & IEC 60318-4 & 1.94 & Electrostatic & {Baseband Displacement} & - \\
			Zheng~\cite{174_zheng_high-spl_2025} & 104 & IEC 60318-4 & 2.1 & Piezoelectric & {Baseband Displacement} & 75 \\
			Chen~\cite{178_chen_design_2025} & 62 & IEC 60318-4 & 2.62 & Piezoelectric & {Baseband Displacement} & - \\
			Lin~\cite{8_lin_bandwidth_2024} & 77 & IEC 60318-4 & 3 & Piezoelectric & {Baseband Displacement} & - \\
			Wang~\cite{9_wang_unsealed_2024} & 104 & IEC 60318-4 & 3.8 & Piezoelectric & {Baseband Displacement} & - \\
			Liechti~\cite{136_liechti_polymer-based_2025} & 106 & IEC 60318-4 & 4.32 & Piezoelectric & {Baseband Displacement} & - \\
			Liechti~\cite{72_liechti_high_2023} & 153 & ff (\SI{1}{\cm}) & 4.84 & Piezoelectric & {Baseband Displacement} & 96.5 \\
			Wang~\cite{179_wang_micropatterned_2025} & 81 & IEC 60318-4 & 5.45 & Piezoelectric & {Baseband Displacement} & - \\
			Kaiser~\cite{68_kaiser_concept_2019} & 70 & IEC 60318-4 & 6.2 & Electrostatic & {Baseband Displacement} & - \\
			Hirano~\cite{144_hirano_pzt_2022} & 98 & IEC 60318-4 & 8.07 & Piezoelectric & {Baseband Displacement} & - \\
			Kim~\cite{154_kim_high_2009} & 128 & ff (\SI{1}{\cm}) & 10.24 & Piezoelectric & {Baseband Displacement} & - \\
			Gao~\cite{148_gao_study_2014} & 136 & ff (\SI{3.16}{\cm}) & 17.35 & Piezoelectric & {Baseband Displacement} & -		
		\end{tabular}
	\end{ruledtabular}
\end{table*}

\end{document}

%% file: schematic_DSR.pdf_tex
%% Creator: Inkscape 1.4 (86a8ad7, 2024-10-11), www.inkscape.org
%% Accompanies image file 'schematic_DSR.pdf' (pdf, eps, ps)
\begingroup%
  \makeatletter%
  \providecommand\color[2][]{%
    \errmessage{(Inkscape) Color is used for the text in Inkscape, but the package 'color.sty' is not loaded}%
    \renewcommand\color[2][]{}%
  }%
  \providecommand\transparent[1]{%
    \errmessage{(Inkscape) Transparency is used (non-zero) for the text in Inkscape, but the package 'transparent.sty' is not loaded}%
    \renewcommand\transparent[1]{}%
  }%
  \providecommand\rotatebox[2]{#2}%
  \newcommand*\fsize{\dimexpr\f@size pt\relax}%
  \newcommand*\lineheight[1]{\fontsize{\fsize}{#1\fsize}\selectfont}%
  \ifx\svgwidth\undefined%
    \setlength{\unitlength}{141.73227265bp}%
    \ifx\svgscale\undefined%
      \relax%
    \else%
      \setlength{\unitlength}{\unitlength * \real{\svgscale}}%
    \fi%
  \else%
    \setlength{\unitlength}{\svgwidth}%
  \fi%
  \global\let\svgwidth\undefined%
  \global\let\svgscale\undefined%
  \makeatother%
  \begin{picture}(1,0.63404405)%
    \lineheight{1}%
    \setlength\tabcolsep{0pt}%
    \put(0,0){\includegraphics[width=\unitlength,page=1]{schematic_DSR.pdf}}%
    \put(0.62756581,0.00367727){\color[rgb]{0,0,0}\makebox(0,0)[lt]{\lineheight{1.25}\smash{\begin{tabular}[t]{l}$t$\end{tabular}}}}%
    \put(-0.00127108,0.50714545){\color[rgb]{0,0,0}\makebox(0,0)[lt]{\lineheight{1.25}\smash{\begin{tabular}[t]{l}$p(t)$\end{tabular}}}}%
    \put(0,0){\includegraphics[width=\unitlength,page=2]{schematic_DSR.pdf}}%
    \put(0.9520976,0.00367727){\color[rgb]{0,0,0}\makebox(0,0)[lt]{\lineheight{1.25}\smash{\begin{tabular}[t]{l}$t$\end{tabular}}}}%
    \put(0,0){\includegraphics[width=\unitlength,page=3]{schematic_DSR.pdf}}%
    \put(0.70633925,0.2520565){\color[rgb]{0,0,0}\makebox(0,0)[lt]{\lineheight{1.25}\smash{\begin{tabular}[t]{l}$p(t)$\end{tabular}}}}%
    \put(0,0){\includegraphics[width=\unitlength,page=4]{schematic_DSR.pdf}}%
    \put(0.18893459,0.60119775){\color[rgb]{0,0,0}\makebox(0,0)[lt]{\lineheight{1.25}\smash{\begin{tabular}[t]{l}inactive speaklet\end{tabular}}}}%
    \put(0.19001401,0.53429125){\color[rgb]{0,0,0}\makebox(0,0)[lt]{\lineheight{1.25}\smash{\begin{tabular}[t]{l}active speaklet\end{tabular}}}}%
    \put(0,0){\includegraphics[width=\unitlength,page=5]{schematic_DSR.pdf}}%
    \put(0.8478456,0.44234745){\color[rgb]{0,0,0}\makebox(0,0)[t]{\lineheight{1.25}\smash{\begin{tabular}[t]{c}ideal\\sound pulse\end{tabular}}}}%
  \end{picture}%
\endgroup%

%% file: mass-spring-oscillator.pdf_tex
%% Creator: Inkscape 1.4 (86a8ad7, 2024-10-11), www.inkscape.org
%% Accompanies image file 'mass-spring-oscillator.pdf' (pdf, eps, ps)
\begingroup%
  \makeatletter%
  \providecommand\color[2][]{%
    \errmessage{(Inkscape) Color is used for the text in Inkscape, but the package 'color.sty' is not loaded}%
    \renewcommand\color[2][]{}%
  }%
  \providecommand\transparent[1]{%
    \errmessage{(Inkscape) Transparency is used (non-zero) for the text in Inkscape, but the package 'transparent.sty' is not loaded}%
    \renewcommand\transparent[1]{}%
  }%
  \providecommand\rotatebox[2]{#2}%
  \newcommand*\fsize{\dimexpr\f@size pt\relax}%
  \newcommand*\lineheight[1]{\fontsize{\fsize}{#1\fsize}\selectfont}%
  \ifx\svgwidth\undefined%
    \setlength{\unitlength}{165.90868252bp}%
    \ifx\svgscale\undefined%
      \relax%
    \else%
      \setlength{\unitlength}{\unitlength * \real{\svgscale}}%
    \fi%
  \else%
    \setlength{\unitlength}{\svgwidth}%
  \fi%
  \global\let\svgwidth\undefined%
  \global\let\svgscale\undefined%
  \makeatother%
  \begin{picture}(1,0.89092869)%
    \lineheight{1}%
    \setlength\tabcolsep{0pt}%
    \put(0,0){\includegraphics[width=\unitlength,page=1]{mass-spring-oscillator.pdf}}%
    \put(0.50786344,0.81960599){\color[rgb]{0,0,0}\makebox(0,0)[rt]{\lineheight{1.25}\smash{\begin{tabular}[t]{r}$F$\end{tabular}}}}%
    \put(0.06394047,0.41036466){\color[rgb]{0,0,0}\makebox(0,0)[rt]{\lineheight{1.25}\smash{\begin{tabular}[t]{r}$b$\end{tabular}}}}%
    \put(0.7871687,0.41036466){\color[rgb]{0,0,0}\makebox(0,0)[rt]{\lineheight{1.25}\smash{\begin{tabular}[t]{r}$k$\end{tabular}}}}%
    \put(0,0){\includegraphics[width=\unitlength,page=2]{mass-spring-oscillator.pdf}}%
    \put(0.40383078,0.72016912){\color[rgb]{0,0,0}\makebox(0,0)[lt]{\lineheight{1.25}\smash{\begin{tabular}[t]{l}$m$\end{tabular}}}}%
    \put(0,0){\includegraphics[width=\unitlength,page=3]{mass-spring-oscillator.pdf}}%
    \put(1.0010043,0.19648675){\color[rgb]{0,0,0}\makebox(0,0)[rt]{\lineheight{1.25}\smash{\begin{tabular}[t]{r}$x$\end{tabular}}}}%
    \put(0,0){\includegraphics[width=\unitlength,page=4]{mass-spring-oscillator.pdf}}%
  \end{picture}%
\endgroup%

%% file: lorentz_force.pdf_tex
%% Creator: Inkscape 1.4 (86a8ad7, 2024-10-11), www.inkscape.org
%% Accompanies image file 'lorentz_force.pdf' (pdf, eps, ps)
\begingroup%
  \makeatletter%
  \providecommand\color[2][]{%
    \errmessage{(Inkscape) Color is used for the text in Inkscape, but the package 'color.sty' is not loaded}%
    \renewcommand\color[2][]{}%
  }%
  \providecommand\transparent[1]{%
    \errmessage{(Inkscape) Transparency is used (non-zero) for the text in Inkscape, but the package 'transparent.sty' is not loaded}%
    \renewcommand\transparent[1]{}%
  }%
  \providecommand\rotatebox[2]{#2}%
  \newcommand*\fsize{\dimexpr\f@size pt\relax}%
  \newcommand*\lineheight[1]{\fontsize{\fsize}{#1\fsize}\selectfont}%
  \ifx\svgwidth\undefined%
    \setlength{\unitlength}{127.55903349bp}%
    \ifx\svgscale\undefined%
      \relax%
    \else%
      \setlength{\unitlength}{\unitlength * \real{\svgscale}}%
    \fi%
  \else%
    \setlength{\unitlength}{\svgwidth}%
  \fi%
  \global\let\svgwidth\undefined%
  \global\let\svgscale\undefined%
  \makeatother%
  \begin{picture}(1,0.67333309)%
    \lineheight{1}%
    \setlength\tabcolsep{0pt}%
    \put(0,0){\includegraphics[width=\unitlength,page=1]{lorentz_force.pdf}}%
    \put(0.10279731,0.38177437){\color[rgb]{0,0,0}\makebox(0,0)[t]{\lineheight{1.25}\smash{\begin{tabular}[t]{c}$i$\end{tabular}}}}%
    \put(0.69887157,0.06403411){\color[rgb]{0,0,0}\makebox(0,0)[rt]{\lineheight{1.25}\smash{\begin{tabular}[t]{r}$\vec{B}$\end{tabular}}}}%
    \put(0,0){\includegraphics[width=\unitlength,page=2]{lorentz_force.pdf}}%
    \put(0.50000023,0.19350319){\color[rgb]{0,0,0}\makebox(0,0)[t]{\lineheight{1.25}\smash{\begin{tabular}[t]{c}$\vec{l}$\end{tabular}}}}%
    \put(0,0){\includegraphics[width=\unitlength,page=3]{lorentz_force.pdf}}%
    \put(0.52329023,0.41867358){\color[rgb]{0,0,0}\makebox(0,0)[lt]{\lineheight{1.25}\smash{\begin{tabular}[t]{l}$\vec{F}_\mathrm{mag}$\end{tabular}}}}%
  \end{picture}%
\endgroup%

%% file: piezoelectric_drive.pdf_tex
%% Creator: Inkscape 1.4 (86a8ad7, 2024-10-11), www.inkscape.org
%% Accompanies image file 'piezoelectric_drive.pdf' (pdf, eps, ps)
\begingroup%
  \makeatletter%
  \providecommand\color[2][]{%
    \errmessage{(Inkscape) Color is used for the text in Inkscape, but the package 'color.sty' is not loaded}%
    \renewcommand\color[2][]{}%
  }%
  \providecommand\transparent[1]{%
    \errmessage{(Inkscape) Transparency is used (non-zero) for the text in Inkscape, but the package 'transparent.sty' is not loaded}%
    \renewcommand\transparent[1]{}%
  }%
  \providecommand\rotatebox[2]{#2}%
  \newcommand*\fsize{\dimexpr\f@size pt\relax}%
  \newcommand*\lineheight[1]{\fontsize{\fsize}{#1\fsize}\selectfont}%
  \ifx\svgwidth\undefined%
    \setlength{\unitlength}{227.19146969bp}%
    \ifx\svgscale\undefined%
      \relax%
    \else%
      \setlength{\unitlength}{\unitlength * \real{\svgscale}}%
    \fi%
  \else%
    \setlength{\unitlength}{\svgwidth}%
  \fi%
  \global\let\svgwidth\undefined%
  \global\let\svgscale\undefined%
  \makeatother%
  \begin{picture}(1,0.24953792)%
    \lineheight{1}%
    \setlength\tabcolsep{0pt}%
    \put(0,0){\includegraphics[width=\unitlength,page=1]{piezoelectric_drive.pdf}}%
    \put(0.1247689,0.10556532){\color[rgb]{0,0,0}\makebox(0,0)[rt]{\lineheight{1.25}\smash{\begin{tabular}[t]{r}$U_{\mathrm{AC}}$\end{tabular}}}}%
    \put(0,0){\includegraphics[width=\unitlength,page=2]{piezoelectric_drive.pdf}}%
    \put(0.59424607,0.11892828){\color[rgb]{0.6627451,0.6627451,0.6627451}\makebox(0,0)[lt]{\lineheight{1.25}\smash{\begin{tabular}[t]{l}$E$\end{tabular}}}}%
    \put(0.95595704,0.13244905){\color[rgb]{0,0,0}\makebox(0,0)[t]{\lineheight{1.25}\smash{\begin{tabular}[t]{c}$e$\end{tabular}}}}%
    \put(0.22386584,0.13244905){\color[rgb]{0,0,0}\makebox(0,0)[t]{\lineheight{1.25}\smash{\begin{tabular}[t]{c}$e$\end{tabular}}}}%
  \end{picture}%
\endgroup%

%% file: pull-configuration.pdf_tex
%% Creator: Inkscape 1.4 (86a8ad7, 2024-10-11), www.inkscape.org
%% Accompanies image file 'pull-configuration.pdf' (pdf, eps, ps)
\begingroup%
  \makeatletter%
  \providecommand\color[2][]{%
    \errmessage{(Inkscape) Color is used for the text in Inkscape, but the package 'color.sty' is not loaded}%
    \renewcommand\color[2][]{}%
  }%
  \providecommand\transparent[1]{%
    \errmessage{(Inkscape) Transparency is used (non-zero) for the text in Inkscape, but the package 'transparent.sty' is not loaded}%
    \renewcommand\transparent[1]{}%
  }%
  \providecommand\rotatebox[2]{#2}%
  \newcommand*\fsize{\dimexpr\f@size pt\relax}%
  \newcommand*\lineheight[1]{\fontsize{\fsize}{#1\fsize}\selectfont}%
  \ifx\svgwidth\undefined%
    \setlength{\unitlength}{204.29950864bp}%
    \ifx\svgscale\undefined%
      \relax%
    \else%
      \setlength{\unitlength}{\unitlength * \real{\svgscale}}%
    \fi%
  \else%
    \setlength{\unitlength}{\svgwidth}%
  \fi%
  \global\let\svgwidth\undefined%
  \global\let\svgscale\undefined%
  \makeatother%
  \begin{picture}(1,0.34783504)%
    \lineheight{1}%
    \setlength\tabcolsep{0pt}%
    \put(0,0){\includegraphics[width=\unitlength,page=1]{pull-configuration.pdf}}%
    \put(0.79187586,0.20666908){\color[rgb]{0,0,0}\makebox(0,0)[rt]{\lineheight{1.25}\smash{\begin{tabular}[t]{r}$F_{\mathrm{el}}$\end{tabular}}}}%
    \put(0.13874939,0.19328004){\color[rgb]{0,0,0}\makebox(0,0)[rt]{\lineheight{1.25}\smash{\begin{tabular}[t]{r}$U_{\mathrm{AC}}$\end{tabular}}}}%
    \put(0,0){\includegraphics[width=\unitlength,page=2]{pull-configuration.pdf}}%
  \end{picture}%
\endgroup%

%% file: push-pull-configuration.pdf_tex
%% Creator: Inkscape 1.4 (86a8ad7, 2024-10-11), www.inkscape.org
%% Accompanies image file 'push-pull-configuration.pdf' (pdf, eps, ps)
\begingroup%
  \makeatletter%
  \providecommand\color[2][]{%
    \errmessage{(Inkscape) Color is used for the text in Inkscape, but the package 'color.sty' is not loaded}%
    \renewcommand\color[2][]{}%
  }%
  \providecommand\transparent[1]{%
    \errmessage{(Inkscape) Transparency is used (non-zero) for the text in Inkscape, but the package 'transparent.sty' is not loaded}%
    \renewcommand\transparent[1]{}%
  }%
  \providecommand\rotatebox[2]{#2}%
  \newcommand*\fsize{\dimexpr\f@size pt\relax}%
  \newcommand*\lineheight[1]{\fontsize{\fsize}{#1\fsize}\selectfont}%
  \ifx\svgwidth\undefined%
    \setlength{\unitlength}{212.80284011bp}%
    \ifx\svgscale\undefined%
      \relax%
    \else%
      \setlength{\unitlength}{\unitlength * \real{\svgscale}}%
    \fi%
  \else%
    \setlength{\unitlength}{\svgwidth}%
  \fi%
  \global\let\svgwidth\undefined%
  \global\let\svgscale\undefined%
  \makeatother%
  \begin{picture}(1,0.60126946)%
    \lineheight{1}%
    \setlength\tabcolsep{0pt}%
    \put(0,0){\includegraphics[width=\unitlength,page=1]{push-pull-configuration.pdf}}%
    \put(0.13320548,0.48387568){\color[rgb]{0,0,0}\makebox(0,0)[rt]{\lineheight{1.25}\smash{\begin{tabular}[t]{r}$-U_{\mathrm{DC}}$\end{tabular}}}}%
    \put(0,0){\includegraphics[width=\unitlength,page=2]{push-pull-configuration.pdf}}%
    \put(0.13416746,0.08211395){\color[rgb]{0,0,0}\makebox(0,0)[rt]{\lineheight{1.25}\smash{\begin{tabular}[t]{r}$+U_{\mathrm{DC}}$\end{tabular}}}}%
    \put(0.13320545,0.28406776){\color[rgb]{0,0,0}\makebox(0,0)[rt]{\lineheight{1.25}\smash{\begin{tabular}[t]{r}$U_{\mathrm{AC}}$\end{tabular}}}}%
    \put(0,0){\includegraphics[width=\unitlength,page=3]{push-pull-configuration.pdf}}%
  \end{picture}%
\endgroup%

%% file: schematic_electrodynamic_speaker.pdf_tex
%% Creator: Inkscape 1.4 (86a8ad7, 2024-10-11), www.inkscape.org
%% Accompanies image file 'schematic_electrodynamic_speaker.pdf' (pdf, eps, ps)
\begingroup%
  \makeatletter%
  \providecommand\color[2][]{%
    \errmessage{(Inkscape) Color is used for the text in Inkscape, but the package 'color.sty' is not loaded}%
    \renewcommand\color[2][]{}%
  }%
  \providecommand\transparent[1]{%
    \errmessage{(Inkscape) Transparency is used (non-zero) for the text in Inkscape, but the package 'transparent.sty' is not loaded}%
    \renewcommand\transparent[1]{}%
  }%
  \providecommand\rotatebox[2]{#2}%
  \newcommand*\fsize{\dimexpr\f@size pt\relax}%
  \newcommand*\lineheight[1]{\fontsize{\fsize}{#1\fsize}\selectfont}%
  \ifx\svgwidth\undefined%
    \setlength{\unitlength}{198.42521848bp}%
    \ifx\svgscale\undefined%
      \relax%
    \else%
      \setlength{\unitlength}{\unitlength * \real{\svgscale}}%
    \fi%
  \else%
    \setlength{\unitlength}{\svgwidth}%
  \fi%
  \global\let\svgwidth\undefined%
  \global\let\svgscale\undefined%
  \makeatother%
  \begin{picture}(1,0.62941043)%
    \lineheight{1}%
    \setlength\tabcolsep{0pt}%
    \put(0,0){\includegraphics[width=\unitlength,page=1]{schematic_electrodynamic_speaker.pdf}}%
    \put(0.49957701,0.27226755){\color[rgb]{0,0,0}\makebox(0,0)[t]{\lineheight{1.25}\smash{\begin{tabular}[t]{c}permanent magnet\end{tabular}}}}%
    \put(0.50020394,0.43334288){\color[rgb]{0,0,0}\makebox(0,0)[t]{\lineheight{1.25}\smash{\begin{tabular}[t]{c}diaphragm\end{tabular}}}}%
    \put(0.49994075,0.00083908){\color[rgb]{0,0,0}\makebox(0,0)[t]{\lineheight{1.25}\smash{\begin{tabular}[t]{c}acoustic holes\end{tabular}}}}%
    \put(0.49978217,0.58702096){\color[rgb]{0,0,0}\makebox(0,0)[t]{\lineheight{1.25}\smash{\begin{tabular}[t]{c}soft magnet\end{tabular}}}}%
    \put(0.50026263,0.50083899){\color[rgb]{0,0,0}\makebox(0,0)[t]{\lineheight{1.25}\smash{\begin{tabular}[t]{c}coil\end{tabular}}}}%
    \put(0,0){\includegraphics[width=\unitlength,page=2]{schematic_electrodynamic_speaker.pdf}}%
  \end{picture}%
\endgroup%

%% file: piezoelectric_drive_d33.pdf_tex
%% Creator: Inkscape 1.4 (86a8ad7, 2024-10-11), www.inkscape.org
%% Accompanies image file 'piezoelectric_drive_d33.pdf' (pdf, eps, ps)
\begingroup%
  \makeatletter%
  \providecommand\color[2][]{%
    \errmessage{(Inkscape) Color is used for the text in Inkscape, but the package 'color.sty' is not loaded}%
    \renewcommand\color[2][]{}%
  }%
  \providecommand\transparent[1]{%
    \errmessage{(Inkscape) Transparency is used (non-zero) for the text in Inkscape, but the package 'transparent.sty' is not loaded}%
    \renewcommand\transparent[1]{}%
  }%
  \providecommand\rotatebox[2]{#2}%
  \newcommand*\fsize{\dimexpr\f@size pt\relax}%
  \newcommand*\lineheight[1]{\fontsize{\fsize}{#1\fsize}\selectfont}%
  \ifx\svgwidth\undefined%
    \setlength{\unitlength}{187.51618393bp}%
    \ifx\svgscale\undefined%
      \relax%
    \else%
      \setlength{\unitlength}{\unitlength * \real{\svgscale}}%
    \fi%
  \else%
    \setlength{\unitlength}{\svgwidth}%
  \fi%
  \global\let\svgwidth\undefined%
  \global\let\svgscale\undefined%
  \makeatother%
  \begin{picture}(1,0.21163556)%
    \lineheight{1}%
    \setlength\tabcolsep{0pt}%
    \put(0,0){\includegraphics[width=\unitlength,page=1]{piezoelectric_drive_d33.pdf}}%
    \put(0.48114216,0.06850762){\color[rgb]{0,0,0}\makebox(0,0)[lt]{\lineheight{1.25}\smash{\begin{tabular}[t]{l}$E$\end{tabular}}}}%
    \put(0.94663832,0.08488916){\color[rgb]{0,0,0}\makebox(0,0)[t]{\lineheight{1.25}\smash{\begin{tabular}[t]{c}$e$\end{tabular}}}}%
    \put(0.05964888,0.08488916){\color[rgb]{0,0,0}\makebox(0,0)[t]{\lineheight{1.25}\smash{\begin{tabular}[t]{c}$e$\end{tabular}}}}%
    \put(0,0){\includegraphics[width=\unitlength,page=2]{piezoelectric_drive_d33.pdf}}%
    \put(0.19778372,0.1741631){\color[rgb]{0,0,0}\makebox(0,0)[t]{\lineheight{1.25}\smash{\begin{tabular}[t]{c}$-$\end{tabular}}}}%
    \put(0.50011971,0.1741631){\color[rgb]{0,0,0}\makebox(0,0)[t]{\lineheight{1.25}\smash{\begin{tabular}[t]{c}$+$\end{tabular}}}}%
    \put(0.8024557,0.1741631){\color[rgb]{0,0,0}\makebox(0,0)[t]{\lineheight{1.25}\smash{\begin{tabular}[t]{c}$-$\end{tabular}}}}%
    \put(0,0){\includegraphics[width=\unitlength,page=3]{piezoelectric_drive_d33.pdf}}%
  \end{picture}%
\endgroup%

%% file: 70_principle.pdf_tex
%% Creator: Inkscape 1.4 (86a8ad7, 2024-10-11), www.inkscape.org
%% Accompanies image file '70_principle.pdf' (pdf, eps, ps)
\begingroup%
  \makeatletter%
  \providecommand\color[2][]{%
    \errmessage{(Inkscape) Color is used for the text in Inkscape, but the package 'color.sty' is not loaded}%
    \renewcommand\color[2][]{}%
  }%
  \providecommand\transparent[1]{%
    \errmessage{(Inkscape) Transparency is used (non-zero) for the text in Inkscape, but the package 'transparent.sty' is not loaded}%
    \renewcommand\transparent[1]{}%
  }%
  \providecommand\rotatebox[2]{#2}%
  \newcommand*\fsize{\dimexpr\f@size pt\relax}%
  \newcommand*\lineheight[1]{\fontsize{\fsize}{#1\fsize}\selectfont}%
  \ifx\svgwidth\undefined%
    \setlength{\unitlength}{141.73234834bp}%
    \ifx\svgscale\undefined%
      \relax%
    \else%
      \setlength{\unitlength}{\unitlength * \real{\svgscale}}%
    \fi%
  \else%
    \setlength{\unitlength}{\svgwidth}%
  \fi%
  \global\let\svgwidth\undefined%
  \global\let\svgscale\undefined%
  \makeatother%
  \begin{picture}(1,0.58198296)%
    \lineheight{1}%
    \setlength\tabcolsep{0pt}%
    \put(0,0){\includegraphics[width=\unitlength,page=1]{70_principle.pdf}}%
    \put(0.50027715,0.52301661){\color[rgb]{0,0,0}\makebox(0,0)[t]{\lineheight{1.25}\smash{\begin{tabular}[t]{c}silicon diaphragm\end{tabular}}}}%
    \put(0,0){\includegraphics[width=\unitlength,page=2]{70_principle.pdf}}%
    \put(0.49887666,0.40198264){\color[rgb]{0,0,0}\makebox(0,0)[t]{\lineheight{1.25}\smash{\begin{tabular}[t]{c}piezoelectric actuator\end{tabular}}}}%
    \put(0.49966919,0.19968598){\color[rgb]{0,0,0}\makebox(0,0)[t]{\lineheight{1.25}\smash{\begin{tabular}[t]{c}rigid\\frame\end{tabular}}}}%
    \put(0,0){\includegraphics[width=\unitlength,page=3]{70_principle.pdf}}%
  \end{picture}%
\endgroup%

%% file: schematic_piezoelectric_spring_speaker.pdf_tex
%% Creator: Inkscape 1.4 (86a8ad7, 2024-10-11), www.inkscape.org
%% Accompanies image file 'schematic_piezoelectric_spring_speaker.pdf' (pdf, eps, ps)
\begingroup%
  \makeatletter%
  \providecommand\color[2][]{%
    \errmessage{(Inkscape) Color is used for the text in Inkscape, but the package 'color.sty' is not loaded}%
    \renewcommand\color[2][]{}%
  }%
  \providecommand\transparent[1]{%
    \errmessage{(Inkscape) Transparency is used (non-zero) for the text in Inkscape, but the package 'transparent.sty' is not loaded}%
    \renewcommand\transparent[1]{}%
  }%
  \providecommand\rotatebox[2]{#2}%
  \newcommand*\fsize{\dimexpr\f@size pt\relax}%
  \newcommand*\lineheight[1]{\fontsize{\fsize}{#1\fsize}\selectfont}%
  \ifx\svgwidth\undefined%
    \setlength{\unitlength}{142.29922557bp}%
    \ifx\svgscale\undefined%
      \relax%
    \else%
      \setlength{\unitlength}{\unitlength * \real{\svgscale}}%
    \fi%
  \else%
    \setlength{\unitlength}{\svgwidth}%
  \fi%
  \global\let\svgwidth\undefined%
  \global\let\svgscale\undefined%
  \makeatother%
  \begin{picture}(1,1.41832635)%
    \lineheight{1}%
    \setlength\tabcolsep{0pt}%
    \put(0,0){\includegraphics[width=\unitlength,page=1]{schematic_piezoelectric_spring_speaker.pdf}}%
    \put(0.65840173,1.30272817){\color[rgb]{0,0,0}\makebox(0,0)[lt]{\lineheight{1.25}\smash{\begin{tabular}[t]{l}diaphragm\end{tabular}}}}%
    \put(0.07984793,1.29661128){\color[rgb]{0,0,0}\makebox(0,0)[lt]{\lineheight{1.25}\smash{\begin{tabular}[t]{l}piezoelectric\\ring actuator\end{tabular}}}}%
    \put(0.67841825,0.36823016){\color[rgb]{0,0,0}\makebox(0,0)[lt]{\lineheight{1.25}\smash{\begin{tabular}[t]{l}springs with\\actuators\end{tabular}}}}%
  \end{picture}%
\endgroup%

%% file: schematic_piezoelectric_speaker.pdf_tex
%% Creator: Inkscape 1.4 (86a8ad7, 2024-10-11), www.inkscape.org
%% Accompanies image file 'schematic_piezoelectric_speaker.pdf' (pdf, eps, ps)
\begingroup%
  \makeatletter%
  \providecommand\color[2][]{%
    \errmessage{(Inkscape) Color is used for the text in Inkscape, but the package 'color.sty' is not loaded}%
    \renewcommand\color[2][]{}%
  }%
  \providecommand\transparent[1]{%
    \errmessage{(Inkscape) Transparency is used (non-zero) for the text in Inkscape, but the package 'transparent.sty' is not loaded}%
    \renewcommand\transparent[1]{}%
  }%
  \providecommand\rotatebox[2]{#2}%
  \newcommand*\fsize{\dimexpr\f@size pt\relax}%
  \newcommand*\lineheight[1]{\fontsize{\fsize}{#1\fsize}\selectfont}%
  \ifx\svgwidth\undefined%
    \setlength{\unitlength}{142.29922557bp}%
    \ifx\svgscale\undefined%
      \relax%
    \else%
      \setlength{\unitlength}{\unitlength * \real{\svgscale}}%
    \fi%
  \else%
    \setlength{\unitlength}{\svgwidth}%
  \fi%
  \global\let\svgwidth\undefined%
  \global\let\svgscale\undefined%
  \makeatother%
  \begin{picture}(1,1.61752967)%
    \lineheight{1}%
    \setlength\tabcolsep{0pt}%
    \put(0,0){\includegraphics[width=\unitlength,page=1]{schematic_piezoelectric_speaker.pdf}}%
    \put(0.10006806,0.53547116){\color[rgb]{0,0,0}\makebox(0,0)[lt]{\lineheight{1.25}\smash{\begin{tabular}[t]{l}piezoelectric layer\end{tabular}}}}%
    \put(0.50003293,0.26983532){\color[rgb]{0,0,0}\makebox(0,0)[t]{\lineheight{1.25}\smash{\begin{tabular}[t]{c}undeflected\end{tabular}}}}%
    \put(0,0){\includegraphics[width=\unitlength,page=2]{schematic_piezoelectric_speaker.pdf}}%
    \put(0.50273948,0.0225645){\color[rgb]{0,0,0}\makebox(0,0)[t]{\lineheight{1.25}\smash{\begin{tabular}[t]{c}deflected\end{tabular}}}}%
    \put(0.87762121,0.5299445){\color[rgb]{0,0,0}\makebox(0,0)[lt]{\lineheight{1.25}\smash{\begin{tabular}[t]{l}slits\end{tabular}}}}%
    \put(0.80061479,1.12520288){\color[rgb]{0,0,0}\makebox(0,0)[t]{\lineheight{1.25}\smash{\begin{tabular}[t]{c}cantilever\\actuator\end{tabular}}}}%
    \put(0.50045286,1.44392734){\color[rgb]{0,0,0}\makebox(0,0)[t]{\lineheight{1.25}\smash{\begin{tabular}[t]{c}rigid\\frame\end{tabular}}}}%
    \put(0,0){\includegraphics[width=\unitlength,page=3]{schematic_piezoelectric_speaker.pdf}}%
  \end{picture}%
\endgroup%

%% file: peripheral_pull-configuration.pdf_tex
%% Creator: Inkscape 1.4 (86a8ad7, 2024-10-11), www.inkscape.org
%% Accompanies image file 'peripheral_pull-configuration.pdf' (pdf, eps, ps)
\begingroup%
  \makeatletter%
  \providecommand\color[2][]{%
    \errmessage{(Inkscape) Color is used for the text in Inkscape, but the package 'color.sty' is not loaded}%
    \renewcommand\color[2][]{}%
  }%
  \providecommand\transparent[1]{%
    \errmessage{(Inkscape) Transparency is used (non-zero) for the text in Inkscape, but the package 'transparent.sty' is not loaded}%
    \renewcommand\transparent[1]{}%
  }%
  \providecommand\rotatebox[2]{#2}%
  \newcommand*\fsize{\dimexpr\f@size pt\relax}%
  \newcommand*\lineheight[1]{\fontsize{\fsize}{#1\fsize}\selectfont}%
  \ifx\svgwidth\undefined%
    \setlength{\unitlength}{204.29950864bp}%
    \ifx\svgscale\undefined%
      \relax%
    \else%
      \setlength{\unitlength}{\unitlength * \real{\svgscale}}%
    \fi%
  \else%
    \setlength{\unitlength}{\svgwidth}%
  \fi%
  \global\let\svgwidth\undefined%
  \global\let\svgscale\undefined%
  \makeatother%
  \begin{picture}(1,0.37733082)%
    \lineheight{1}%
    \setlength\tabcolsep{0pt}%
    \put(0,0){\includegraphics[width=\unitlength,page=1]{peripheral_pull-configuration.pdf}}%
    \put(0.71393391,0.28258789){\color[rgb]{0,0,0}\makebox(0,0)[rt]{\lineheight{1.25}\smash{\begin{tabular}[t]{r}$F_{\mathrm{el}}$\end{tabular}}}}%
    \put(0.13874939,0.24955933){\color[rgb]{0,0,0}\makebox(0,0)[rt]{\lineheight{1.25}\smash{\begin{tabular}[t]{r}$U_{\mathrm{AC}}$\end{tabular}}}}%
    \put(0,0){\includegraphics[width=\unitlength,page=2]{peripheral_pull-configuration.pdf}}%
  \end{picture}%
\endgroup%